\documentclass[11pt,a4paper]{article}
\usepackage{jinstpub}
\usepackage{lineno}
\usepackage{comment}
\usepackage{orcidlink}
\usepackage{color}
\usepackage{siunitx}
\usepackage{subcaption}
\usepackage[colorinlistoftodos,prependcaption,textsize=tiny]{todonotes}
\usepackage{url}

\DeclareSIUnit\bar{bar}

\newcommand{\iselSystem}{isel-system}

\newcommand{\figsubref}[1]{(\protect\subref{#1})}
\newcommand{\figref}[1]{Figure~\ref{#1}}
\newcommand{\figrefbra}[1]{Fig.~\ref{#1}}
\newcommand{\Figref}[1]{Figure~\ref{#1}}

\newcommand{\tabref}[1]{Table~\ref{#1}}
\newcommand{\tabrefbra}[1]{Tab.~\ref{#1}}
\newcommand{\Tabref}[1]{Table~\ref{#1}}

\newcommand{\secref}[1]{Section~\ref{#1}}
\newcommand{\secrefbra}[1]{Sec.~\ref{#1}}
\newcommand{\Secref}[1]{Section~\ref{#1}}

\newcommand{\eqnref}[1]{Eqn.~\eqref{#1}}

\title{GRANITE: Mechanical Characterization and Optical Inspection of Large-Area TPC Electrodes}

\author[a,1]{Alexander~Deisting,\,\orcidlink{0000-0001-5372-9944},\note{Corresponding author}}
\emailAdd{alexander.deisting@cern.ch}
\author[a]{Jan~Lommler\,\orcidlink{0009-0004-9049-2199},}
\author[a]{Shumit~Mitra\,\orcidlink{0009-0008-0610-6152},}
\author[a,b]{Uwe~Oberlack\,\orcidlink{0000-0001-8160-5498},}
\author[b]{Fabian~Piermaier\,\orcidlink{0000-0002-3863-3394},}
\author[b]{Quirin~Weitzel\,\orcidlink{0000-0002-3073-8642},}
\author[a,2]{and Daniel~Wenz\,\orcidlink{0009-0004-5242-3571}\note{Now at: Institut für Kernphysik, Universität Münster, Wilhelm-Klemm-Str. 9, 48149 Münster, Germany}}
\affiliation[a]{Institut für Physik \& Exzellenzcluster PRISMA+, Johannes Gutenberg-Universität Mainz, Staudingerweg 7, 55128 Mainz, Germany}
\affiliation[b]{Detektorlabor, Exzellenzcluster PRISMA$^{+}$, Johannes Gutenberg-Universität Mainz, Staudingerweg 9, 55128 Mainz, Germany}

\abstract{Next-generation dual-phase time projection chambers (TPCs) for rare event searches will require large-scale, high-precision electrodes. To meet the stringent requirements for mechanical stability and high-voltage performance of such an experiment, we have developed a scanning setup for comprehensive electrode quality assurance called GRANITE: Granular Robotic Assay for Novel Integrated TPC Electrodes. GRANITE is built around a gantry robot on top of a $\SI{2.5}{\meter}\times\SI{1.8}{\meter}$ granite table, equipped with a suite of non-contact metrology devices.\\
We demonstrate the setup's capabilities in two key areas: first, using laser scanners, we characterize wire tension, and in an independent measurement wire deflection due to gravity and electrostatic forces is determined. The setup achieves a precision of \SI{20}{\micro\meter} for the relative measurement of only electrostatic displacement. Furthermore, GRANITE can measure gravitational sag down to \SI{200}{\micro\meter} in an absolute measurement; this precision improves to \SI{50}{\micro\meter} when applying model-based corrections for systematic effects. The performance achieved exceeds the needs for the characterisation of the electrode sagging in future experiments, which typically aims to ensure a maximal sag on the order of \SI{500}{\micro\meter}.\\
Second, we use GRANITE's high resolution camera to image every wire of the cathode grid of the XENON1T experiment. Subsets of these images are then hand sorted and used to train an autoencoder, to reliably classify wire images as either pristine wires or images containing severe anomalous features. These anomalies appear \textit{e.g.} as staining and may be potential defects. The interpretation of the classification results is complicated by the fact that most wire segments are not spotless, but show a varying amount of anomalous features. Follow-up studies are needed to identify the exact nature of such features on wires and if they cause effects (\textit{e.g.} field emission) which would prohibit the deployment of the corresponding wire as part of an electrode in a future dual-phase TPC.}


\arxivnumber{2511.11400}

\begin{document}

\maketitle

\section{Introduction}
\label{sec:introducion}

Liquid xenon (LXe) filled dual-phase time projection chambers (TPCs) have provided the most stringent limits on weakly interacting massive particle (WIMP) dark matter (DM) \cite{PhysRevLett.131.041003,4dyc-z8zf,PhysRevLett.134.011805} for WIMP masses $\gtrsim\!\!\SI{5}{\giga\electronvolt\per c^2}$ to date. These detectors are the result of ambitious research and development programmes increasing their size and decreasing their intrinsic background rates. 
Xenon filled dual-phase TPCs have the ability to discriminate between nuclear recoils (NR, \textit{e.g.} caused by a WIMP), and electron recoils (ER, \textit{e.g.} caused by $\gamma$-radiation of trace ambient radioactivity). This feature is key to the excellent background reduction of these detectors. 
Planning for the next generation of xenon filled DM observatory has begun; it will cover a rich science programme of low-background and rare-event searches, \textit{e.g.} searches for $0\nu\beta\beta$ decays of $^{136}\text{Xe}$. Some of these science channels will rely on the detection of events with very little energy deposited inside the detector, \textit{e.g.} coherent elastic neutrino nucleus scattering (CE$\nu$NS). First indications of this process have been seen by current experiments \cite{PhysRevLett.133.191002,PhysRevLett.133.191001} and a future observatory is expected to discover it.\\
{\indent}The XLZD collaboration is proposing a dual-phase TPC with approximately \SI{3}{\meter} diameter and height \cite{XLZD:2024nsu}, where its design has been informed by the preceding detectors, in particular by the XENONnT \cite{XENON:2024wpa} and LZ \cite{AKERIB2020163047} experiments. These TPCs -- and PandaX-4T \cite{PhysRevLett.127.261802} -- have electrodes with \SI{1}{\meter} to \SI{1.5}{\meter} diameter, therefore XLZD aims at a scale more than doubling the current TPCs' dimensions. The main electrodes are cathode, gate electrode, and anode. The gate electrode and the anode sandwich the liquid-gas interface, whilst the cathode is located on the lower end of the TPC. An interaction of a particle will produce primary scintillation light (S1) and electrons, which move in the electric field (``drift field'') between gate and cathode towards the top of the detector. A high electric field between gate and anode (``extraction field'', $E_\text{ext}$) accelerates these electrons, so they have enough energy to cross the liquid-gas interface. There they produce electroluminescence light in the xenon gas -- the S2 signal. S1 and S2 signals are both read out with arrays of light sensors at the top and bottom of the TPC.\\
{\indent}Particles interacting either via ER or NR deposit a different fraction of their energy in primary scintillation light (S1) as well as primary ionisations (S2). The magnitude of the drift field determines the amount of primary ionisations that recombines once more and thus contribute to the S1 signal. In LXe optimal NR vs ER discrimination is achieved for a drift field on the order of a very few \SI{100}{\volt\per\centi\meter} \cite{Szydagis:2022ikv}. Thus, for \SI{1}{\meter} of cathode-to-gate distance, a cathode voltages of at least \SI{10}{\kilo\volt} is needed. 
Asperities or other defects on the electrode material may lead to field-emission of electrons, in particular in the presence of high $E_\text{ext}$. These electrons could mimic low energy signals in the detector, contribute to the accidental coincident background, or -- in the worst case scenario -- they could initiate breakdown between electrodes and render the detector unusable. Various values can be found for the surface field at which electron emission occurs, \textit{e.g.} $\sim\!\!\SI{1E+6}{\volt\per\centi\meter}$ \cite{raizer1997gas} or $\sim\!\!\SI{1E+4}{\volt\per\centi\meter}$ \cite{TOMAS201849}, where the latter number has been observed in a dedicated dual-phase xenon configuration, finding that the general surface treatment of the wire drives current emission, rather than single protrusions. The Malter effect \cite{PhysRev.50.48} causes electron emission from a cathodic surface, such as the gate electrode or the cathode in a dual phase TPC, covered with a thin insulating layer, \textit{e.g.} certain oxides. Positive ions can accumulate on this layer and through their space-charge create a localised electric field high enough to pull electrons from the conductor through the insulating layer into the surrounding medium. This effect is of particular concern for gaseous detectors \cite{Rolandi:2008ujz,raizer1997gas}, and is discussed as a possible reason for electron emission in LXe filled detectors in \cite{Vavra:2025fif}.\\
{\indent}To construct a dual-phase TPC's electrodes a careful inspection of the wires is needed. Wires are unwound, cut, and mounted on a temporary frame. Then they are cleaned, \textit{e.g.} in an ultrasonic bath with alkaline soaps, rinsed, and passivated with a solution of a few percent of citric or nitric acid to allow for an oxidation layer to build up \cite{LINEHAN2022165955,XENON:2024wpa}. One motivation for the passivation processes is to remove the existing oxidation layer from the wires and to form a new layer under controlled conditions. Finally, the wires are stretched and mounted on a frame. A different order of these steps is possible \cite{LINEHAN2022165955,Stifter_2020}. The inspection of wires may happen by optical means, by testing the high voltage (HV) performance, or a combination thereof.\\
{\indent}Testing electrodes of the size needed for XLZD or another dual-phase TPC in a cryogneic liquid is time-consuming. Therefore, such tests (as \textit{e.g.} described in \cite{brown2024pancake}) should be preceded by tests conducted in air or a gas cheaper than xenon. For this reason, we have developed the setup described in this paper. GRANITE, short for Granular Robotic Assay for Novel Integrated TPC Electrodes, is capable of wire assessment via high-resolution images, as well as measuring wire tension, and wire deflection with and without electric fields via laser distance sensors. It allows scans of electrodes in excess of \SI{2}{\meter\squared} area, while utilising methods that are transferable to larger setups.\\[0.2cm]
{\indent}The paper is structured as follows: In \secref{sec:setup} the GRANITE setup is described. \Secref{sec:sagging:test-wires} shows example tension and sagging measurements using laser distance sensors, their results, and discusses how these match the accuracy needed for the assay of future TPC electrodes. Then, we image the full XENON1T cathode grid \cite{XENON:2017lvq} with the high-resolution camera of GRANITE. A fraction of the resulting images are used to train an autoendocer for anomaly detection, classifying images of wire segments either as pristine wire segments or such with visible features, which may be defects. 
Finally, we conclude the full paper in \secref{sec:conclusions:outlook} and give an outlook on additional measurements, which may aid identifying the kinds of anomaly leading to problems inside a detector as \textit{e.g.} field emission of electrons from electrode wire surfaces.

\section{The GRANITE setup}
\label{sec:setup}

\begin{figure}
  \centering
  \subfloat[]{\label{sec:setup:fig:granite_table:schematic}\includegraphics[height=0.18\textheight]{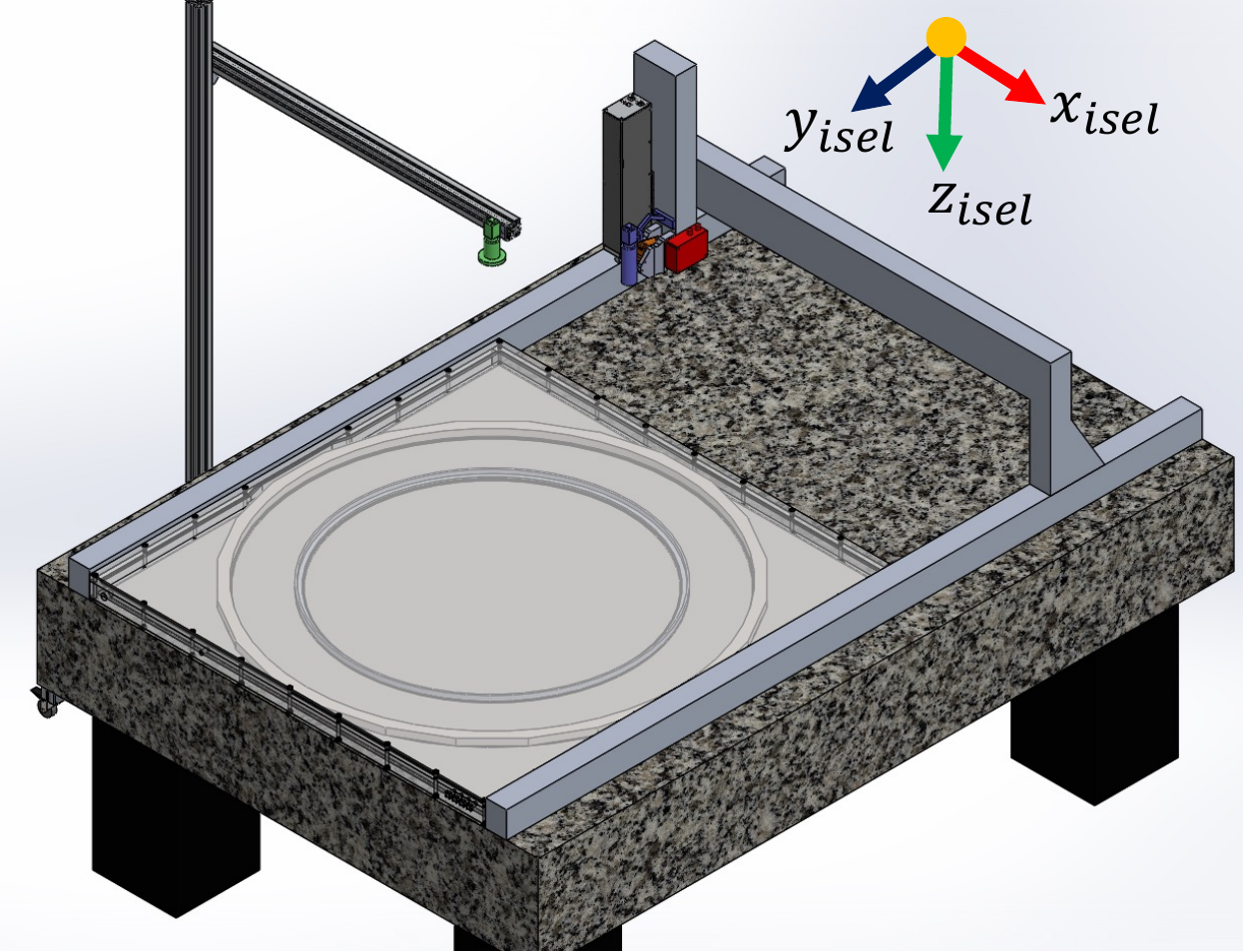}}
  \subfloat[]{\label{sec:setup:fig:granite_table:photo}\includegraphics[height=0.18\textheight]{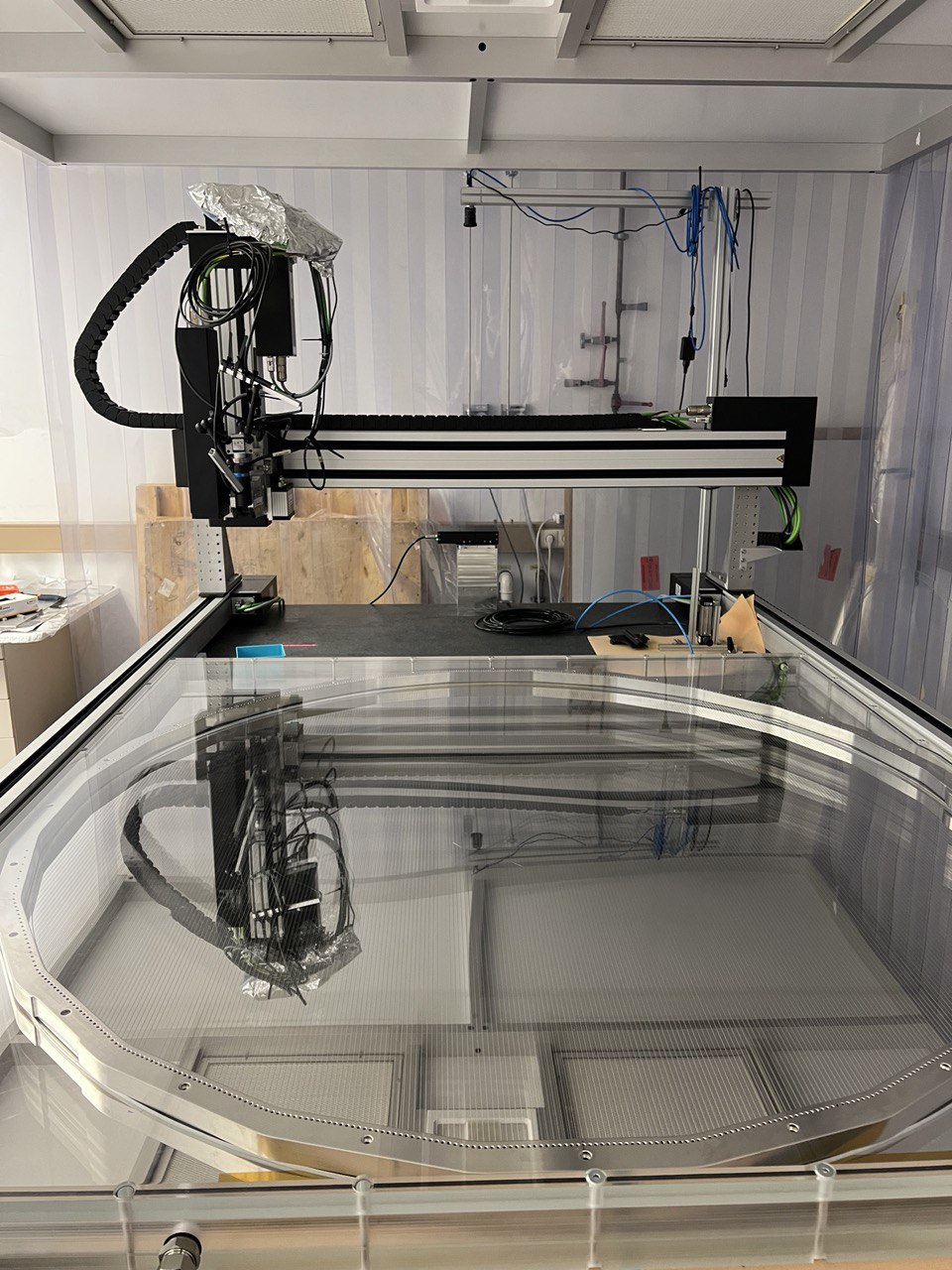}}
  \subfloat[]{\label{sec:setup:fig:granite_table:metrology}\includegraphics[height=0.18\textheight, trim = 0 0 0 0,clip=true]{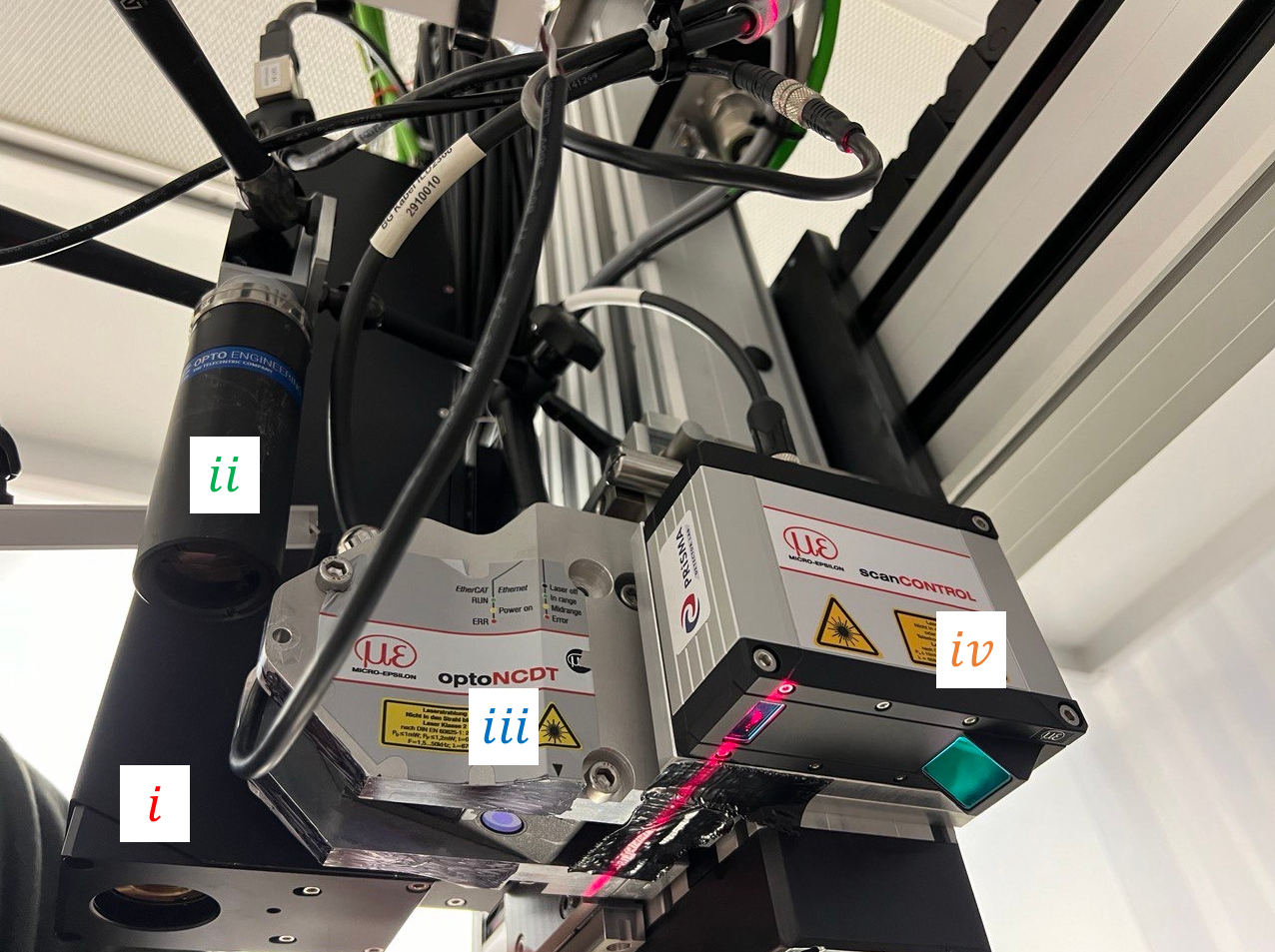}}
    \caption{\label{sec:setup:fig:granite_table}\figsubref{sec:setup:fig:granite_table:schematic} Schematic of the gantry setup on the $2.5 \times \SI{1.8}{\meter\squared}$ granite table with coordinate system. \figsubref{sec:setup:fig:granite_table:photo} Photo of the setup: In the foreground an acrylic glass box with a $\sim\!\!\SI{1.4}{\meter}$ diameter electrode can be seen, whilst the arm with the metrology components described in the text is visible on the far, left side of the table. \figsubref{sec:setup:fig:granite_table:metrology} The different metrology components from left to right: Confocal microscope (\textcolor{red}{$i$}) and -- slightly higher -- the high resolution (industrial) camera (\textcolor{green!45!black}{$ii$}), the laser distance sensor (\textcolor{blue}{$iii$}), and the profile laser scanner (\textcolor{orange}{$iv$}).}
\end{figure}
\begin{figure}[t]
\centering
  \begin{minipage}{0.25\columnwidth}
    \centering
    \subfloat[]{
      \label{sec:setup:fig:opticalSensors:confocal-microscope}
      \includegraphics[trim= 0 0 0 0, clip=true, width = 0.99\columnwidth]{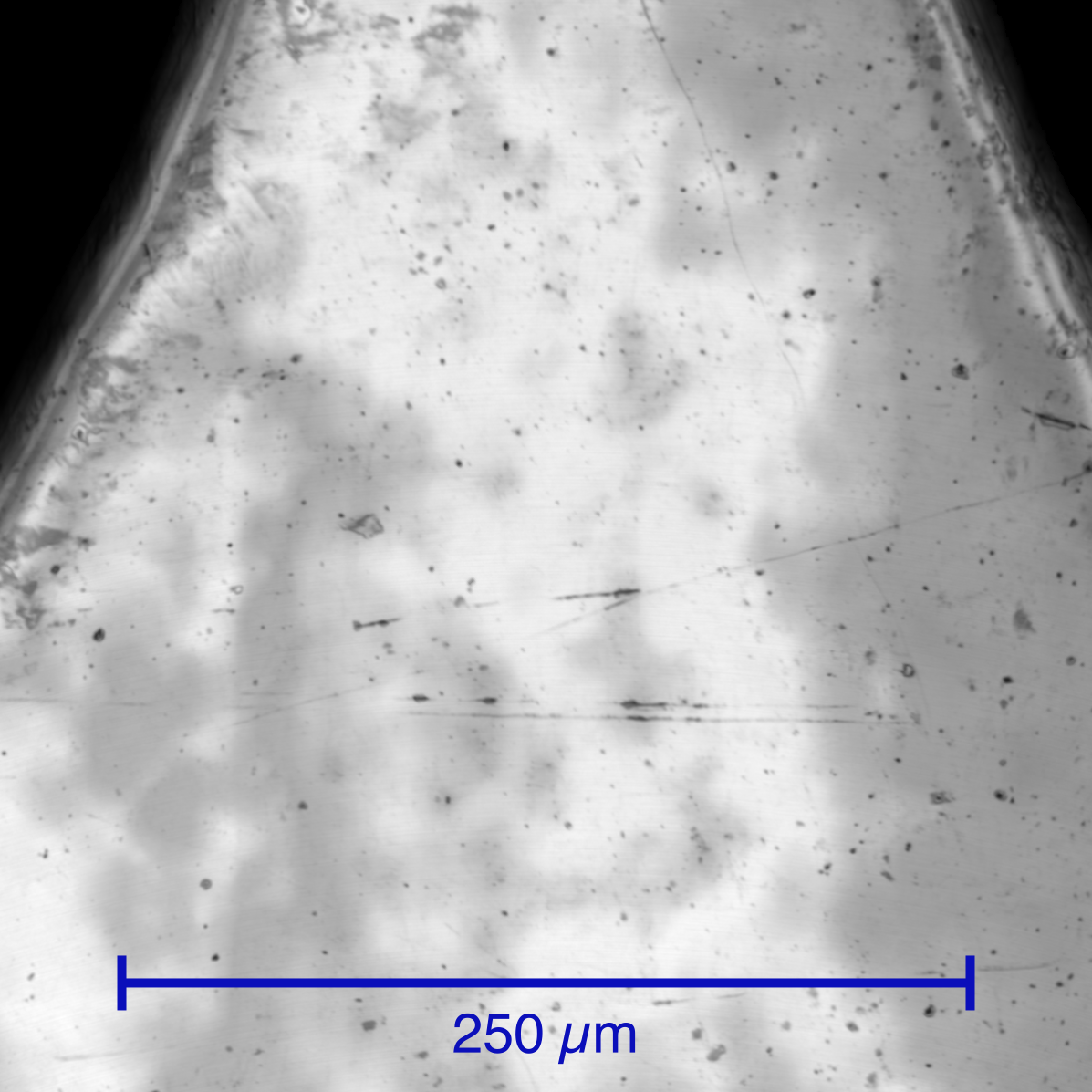}}\\[0.2cm]
    \subfloat[]{
      \label{sec:setup:fig:opticalSensors:high-res-cam}
      \includegraphics[trim= 0 0 0 0, clip=true, width = 0.99\columnwidth]{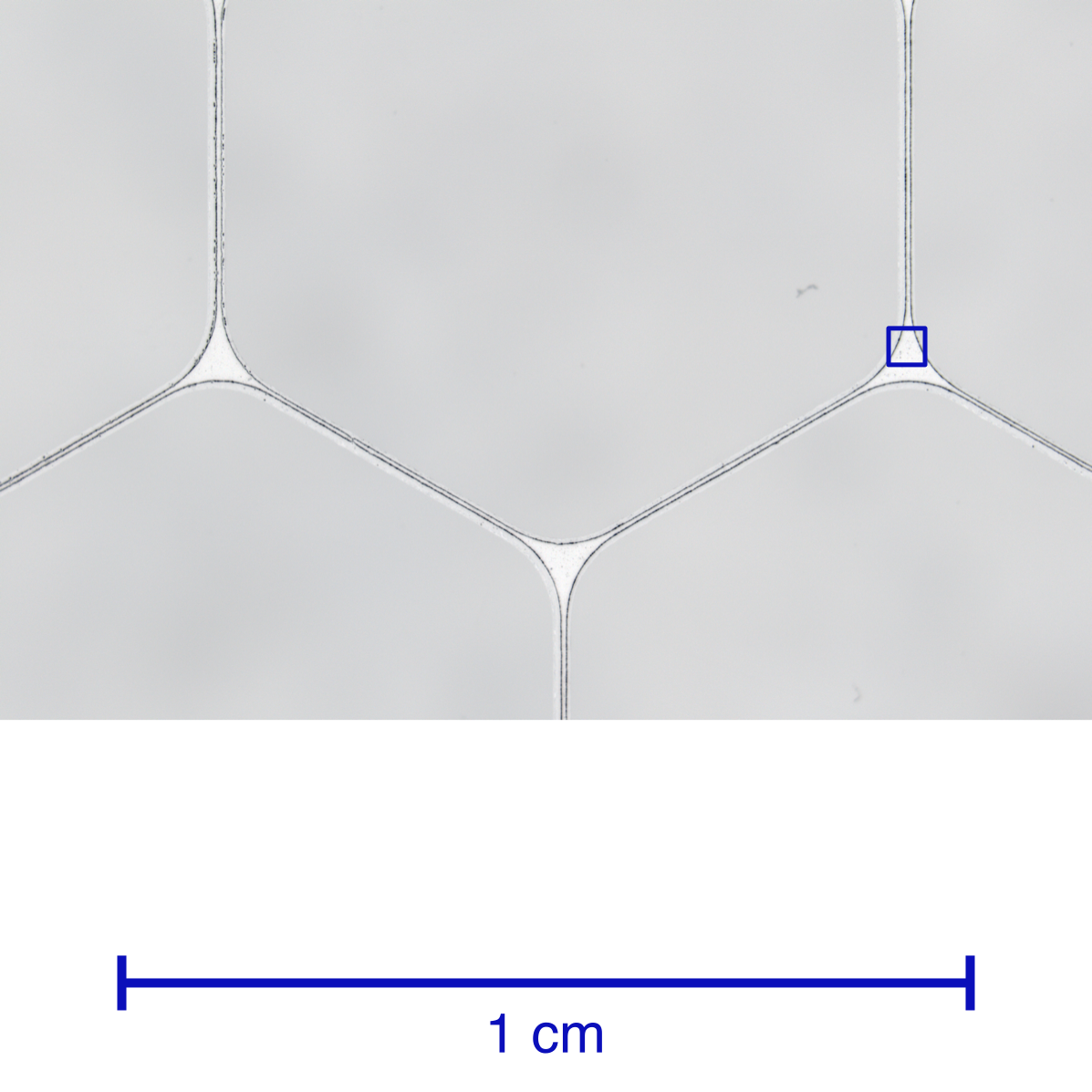}}
  \end{minipage}
  \hspace{0.03\columnwidth}
  \begin{minipage}{0.63\columnwidth}
    \centering
    \subfloat[]{
      \label{sec:setup:subsec:calibration:pointlaser:fig:fulltable:withcuts}
      \includegraphics[width=0.90\columnwidth, trim = 0 0 0 0 0, clip= true]{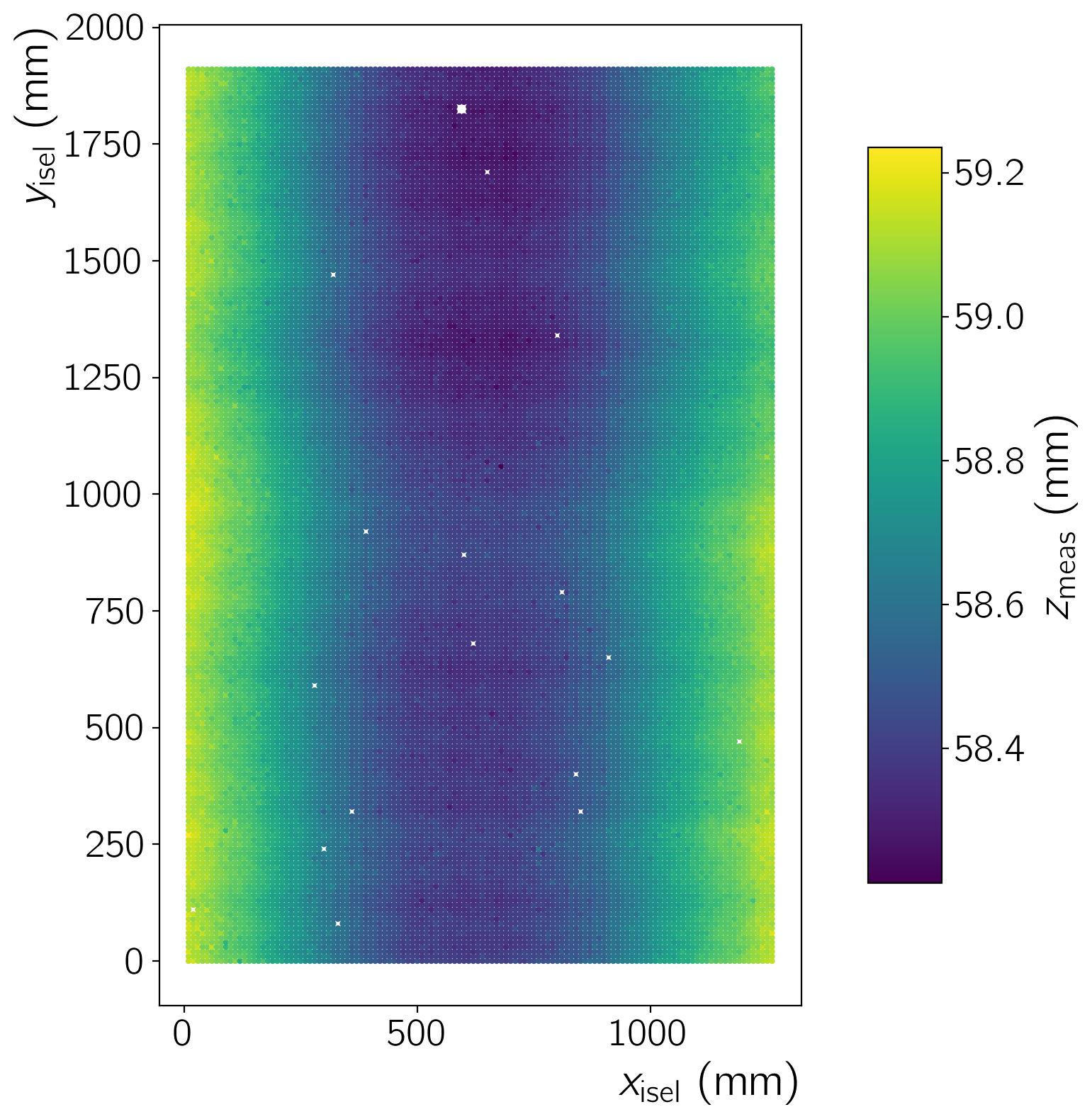}}
  \end{minipage}
  \caption{\label{sec:setup:fig:opticalSensors}Illustration of the resolution of the \figsubref{sec:setup:fig:opticalSensors:confocal-microscope} \textit{confocal microscope} and a $\times50$ lens, and \figsubref{sec:setup:fig:opticalSensors:high-res-cam} the \textit{high resolution industrial camera}. The region marked in blue shows the size of the confocal microscope image. \figsubref{sec:setup:subsec:calibration:pointlaser:fig:fulltable:withcuts} Data obtained with the \textit{optoNCDT 2300-20} laser distance sensor when measuring the distance of the sensor to the empty table. The sagging of the ``gantry bridge'' can be seen as the valley at $x_\text{isel}\!\!\sim\!\!\SI{600}{\milli\meter}$. Oscillations in $z_\text{meas}$ as function of $y_\text{isel}$ are visible as well.}
\end{figure}
The GRANITE setup, shown in \figref{sec:setup:fig:granite_table}, is built on a granite table, providing a smooth and level surface of roughly $2.5 \times \SI{1.8}{\meter\squared}$, as well as vibration damping due to its mass. A laminar flow unit housing the granite table is used to keep the lab environment clean. The whole setup is located in a temperature controlled lab at the PRISMA$^{+}$ excellence cluster of the Johannes Gutenberg University Mainz. Temperature, together with relative humidity, dew-point and atmospheric pressure are recorded. A gantry robot is mounted on top of the table (``\iselSystem{}''), built from parts provided by iselGermany, holding multiple optical measurement devices, visible in \figref{sec:setup:fig:granite_table:metrology}: 
(\textit{i}) a confocal microscope (\textit{NanoFocus µsurf 350 HDR F}, example image is shown in \figrefbra{sec:setup:fig:opticalSensors:confocal-microscope}), 
(\textit{ii}) an industrial camera \cite{baslerweb:acA4600-7gc} (\textit{Basler acA4600-7gc}) with a telecentric lens \cite{opto-e:TC23016} (field depth \SI{1.5}{\milli\meter}, working distance \SI{43.1}{\milli\meter}, example image \figrefbra{sec:setup:fig:opticalSensors:high-res-cam}), 
(\textit{iii}) a laser distance sensor \cite{muepsilon:optoNCDT2300} (\textit{optoNCDT 2300-20}), and 
(\textit{iv}) a profile laser scanner \cite{muepsilon:scanCONTROL3000} (\textit{scanCONTROL 3000}). Each of these devices can be used for a different measurement:
\begin{itemize}
  \setlength\itemsep{-0.1em}
  \item 3D surface imaging of the wire with sub-micron precision (\textit{i}) 
  \item High quality pictures of the wire surface (\textit{ii}) 
  \item Sagging measurement of a single wire or multiple wires simultaneously, calibration measurements of the table and gantry (\textit{iii}), (\textit{iv}) 
  \item Measurements of distance oscillations (\textit{iii}) 
\end{itemize}
{\indent}The measurement devices can be moved with the gantry robot enabling scans of surfaces up to an area of about $2.0 \times \SI{1.4}{\meter\squared}$. 
Laser distance sensor and \iselSystem{} are identical to \cite{8069756}. A repeatability of $\pm\SI{20}{\micro\meter}$ 
is quoted by the producer. The minimum step-size in direction of $x_\text{isel}$ and $y_\text{isel}$ ($z_\text{isel}$) is \SI{5}{\micro\meter} (\SI{2.5}{\micro\meter}).\\
{\indent}A custom acrylic glass enclosure has been made which facilitates the inspection of the wire electrodes under different electric fields in a controlled atmosphere (e.g. argon). Inside the enclosure is an ultra-high polished stainless steel plate for grounding. Electrodes lie on top of acrylic glass spacers. With this configuration, the electrodes can be examined at the desired electric fields. HV is supplied from different CAEN power supply modules \cite{CAEN:N1470,CAEN:DT1471ET}. 
Custom \textsc{python} software controls the movement of the gantry robot, and the data acquisition of the camera and the laser distance sensors, as well as the CAEN power supplies.

\subsection{Gantry Features Visible in the Laser-Distance Measurements}
\label{sec:setup:subsec:gantry:features}

\begin{figure}
  \centering
  \subfloat[]{\label{sec:setup:subsec:gantry:features:fig:explanation:sketch}
    \includegraphics[height=0.21\textheight, trim = 0 0 0 40, clip= true]{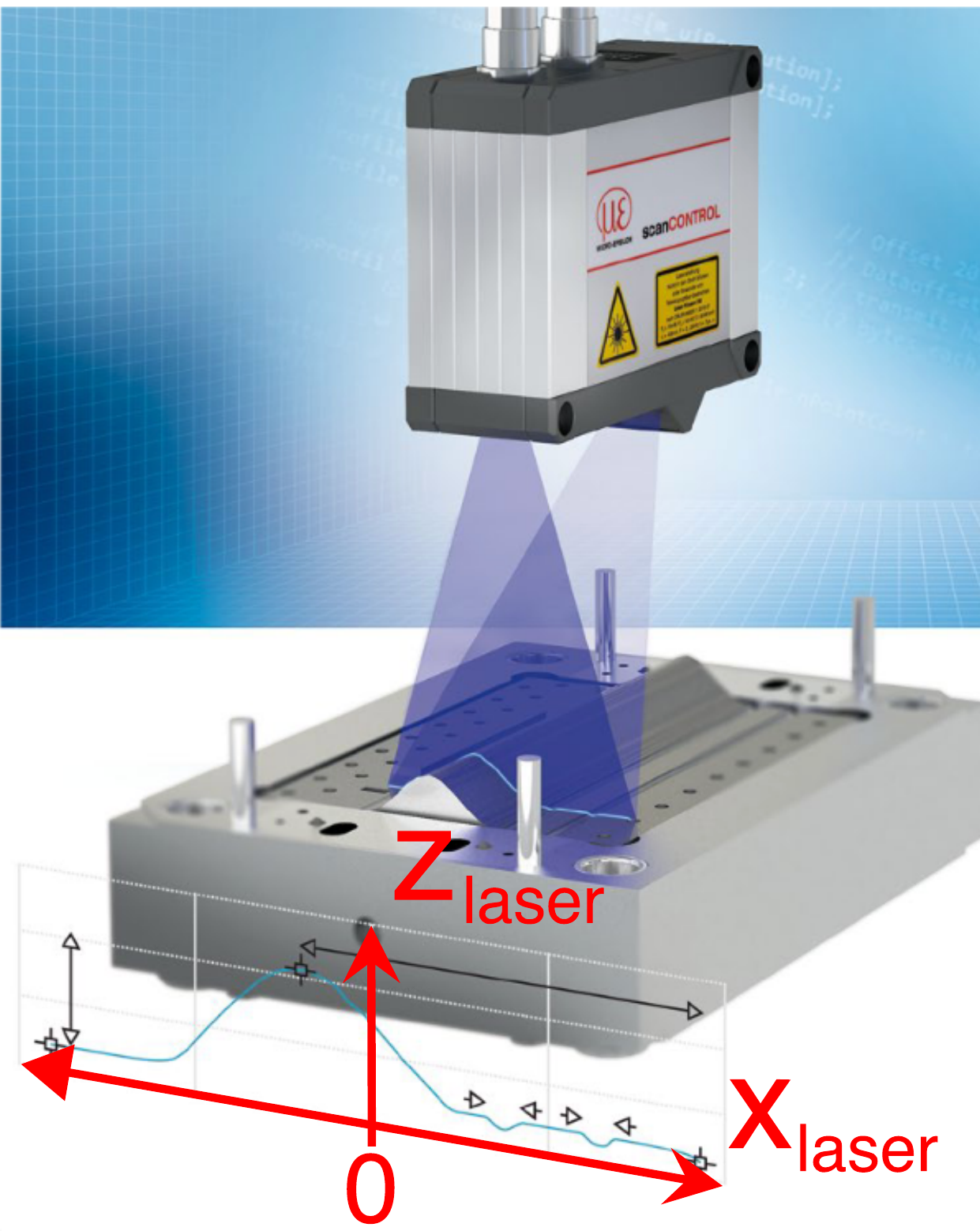}}
  \subfloat[]{\label{sec:setup:subsec:gantry:features:fig:linelaserexplanation:datanoncorr}
    \includegraphics[height=0.21\textheight, trim = 0 0 0 0, clip= true]{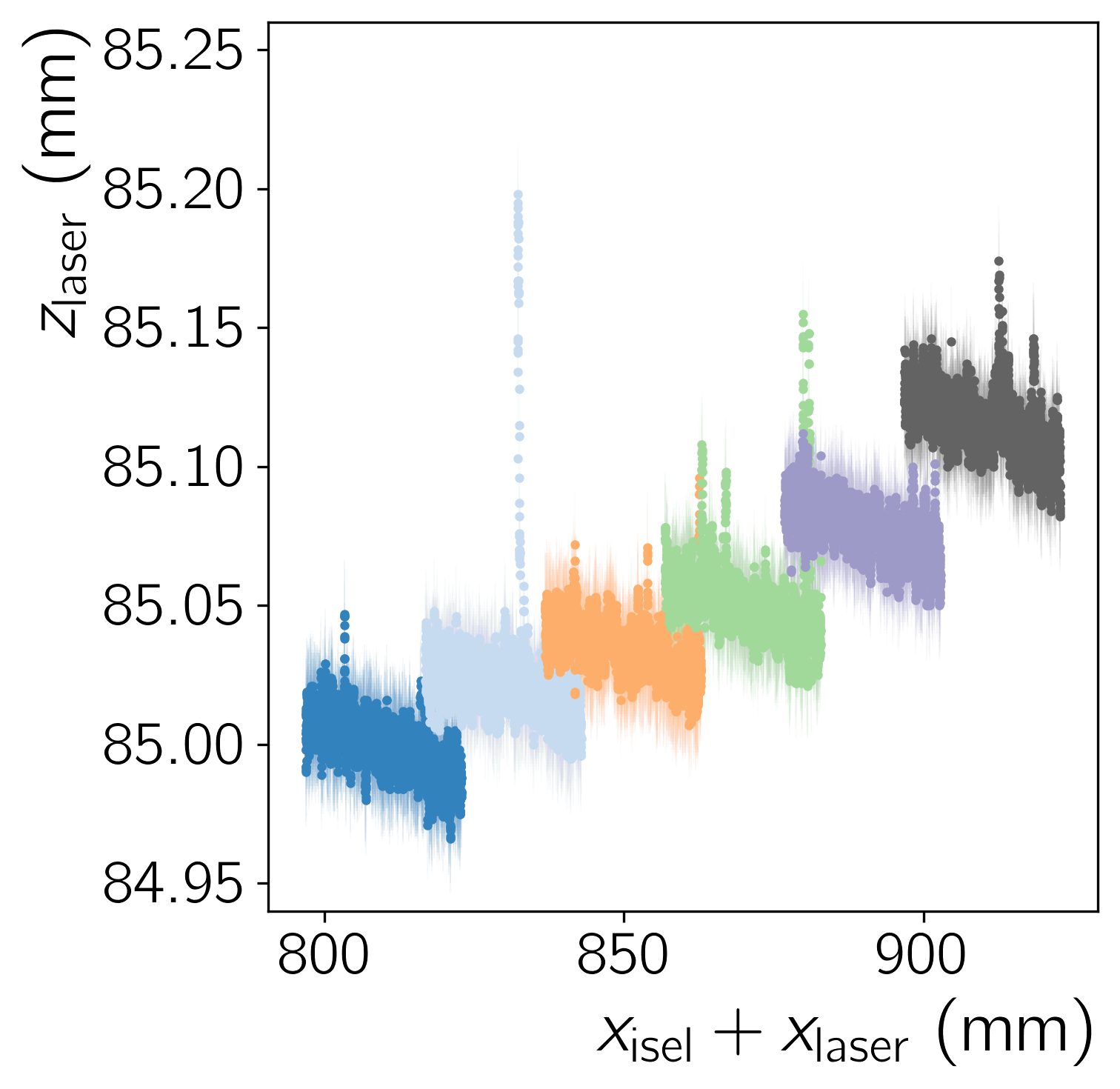}}
  \subfloat[]{\label{sec:setup:subsec:gantry:features:fig:linelaserexplanation:datacorr}
    \includegraphics[height=0.21\textheight, trim = 0 0 0 0, clip= true]{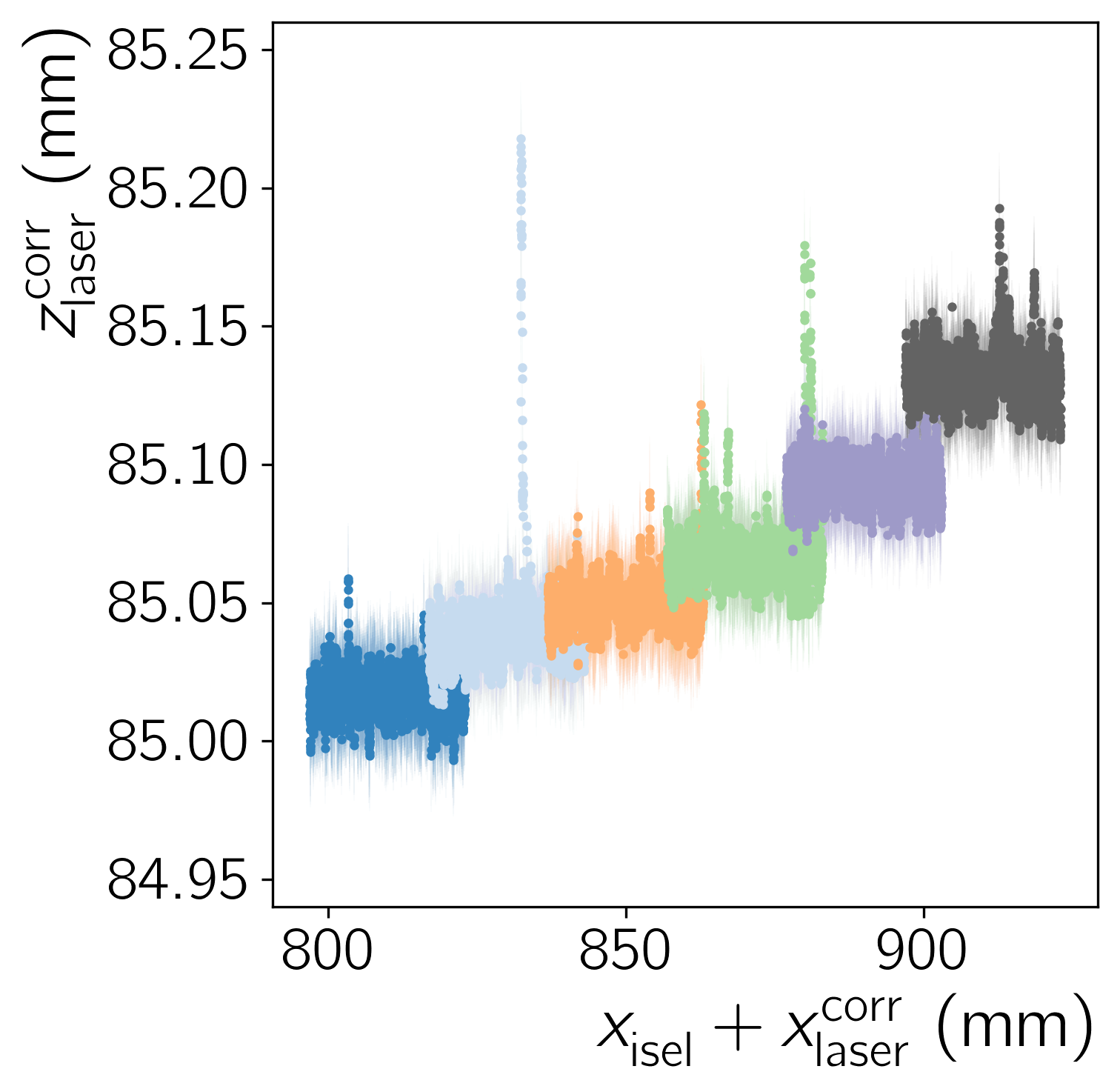}}
  \caption{\label{sec:setup:subsec:calibration:pointlaser:linelaser}\figsubref{sec:setup:subsec:gantry:features:fig:explanation:sketch} A graphic from the data-sheet \cite{muepsilon:scanCONTROL3000} of the \textit{scanCONTROL 3000} profile laser scanner overlaid with the measurement coordinate system used in this paper. \figsubref{sec:setup:subsec:gantry:features:fig:linelaserexplanation:datanoncorr} Six height profiles above the flat table surface measured by the laser scanner ($y_{\text{isel}}=\SI{0}{\milli\meter}$). Colours are used to distinguish the measurements at different $x_{\text{isel}}$ positions. The profile laser scanner's tilt is visible from the left to right in every profile. \figsubref{sec:setup:subsec:gantry:features:fig:linelaserexplanation:datacorr} The same data as in \figsubref{sec:setup:subsec:gantry:features:fig:linelaserexplanation:datanoncorr}, but with the tilt correction applied. The height differences between the profiles are due to the sagging of the gantry bridge, which changes the distance to the granite table for different $x_{\text{isel}}$ positions.}
\end{figure}

\Figref{sec:setup:subsec:calibration:pointlaser:fig:fulltable:withcuts} shows a series of measurement points with the \textit{optoNCDT 2300-20} sensor over the full area of the empty granite table. The table's flatness after polishing was better than \SI{14}{\micro\meter} \cite{8069756} and may have degraded subsequently. 
The figure shows a changing distance between laser sensor and table surface as a function of position above the table, with a variation of more than \SI{800}{\micro\meter} between the lowest and highest point. 
The sagging of the \iselSystem{}'s ``gantry bridge'' gives rise to the valley at $x_\text{isel}\sim\!\!\SI{600}{\milli\meter}$. The gantry is moved via ball screws. Each axis is supported and moved by two such screws, which have zero clearance due to their pre-tension. Nevertheless, an oscillation along the $y_\text{isel}$-axis of an amplitude less than \SI{100}{\micro\meter} can be seen in \figref{sec:setup:subsec:calibration:pointlaser:fig:fulltable:withcuts}. This is likely caused by the way the axis are mounted to the table and/or the finite precision of the manufacturing of the axis.\\[0.2cm]
{\indent}As reported in \cite{8069756}, there is a drift in time of the exact manifestation of the gantry features (bridge sagging, oscillations) affecting the laser-to-table distance. 
The corresponding distance offset depends on the exact $x_\text{isel},y_\text{isel}$ coordinate and on the direction of movement to said coordinates. For these reasons it is not viable to use a measurement as displayed in \figref{sec:setup:subsec:calibration:pointlaser:fig:fulltable:withcuts} as a correction map for future measurements. 
One can minimise the expected impact of these features on a measurement by aligning an object under test in parallel to the $y_\text{isel}$ direction to not be 
affected by the bridge sagging. 
However, the measurement will still show the oscillation along $y_\text{isel}$. 
The highest precision is reached when using relative measurements (\textit{e.g.} of a wire grid with and without voltage applied), as all position- and movement-dependent distance offsets due to the gantry cancel out, provided the measurements are performed in short order. Therefore, our work presented in \secref{sec:sagging:test-wires} relies on relative measurements when possible.\\
{\indent}In case of the \textit{scanCONTROL 3000} profile laser scanner, there is one more ``feature'' to take care of. \Figref{sec:setup:subsec:gantry:features:fig:explanation:sketch} illustrates the way this laser scanner measures one-dimensional height profiles $z_{\text{laser}}\left(x_{\text{laser}}\right)$. There are 2048 measurement points between $-\SI{12.5}{\milli\meter}\leq x_{\text{laser}}\leq\SI{12.5}{\milli\meter}$. 
The profile laser is mounted such that $x_{\text{laser}}$ is parallel to $x_{\text{isel}}$. \Figref{sec:setup:subsec:gantry:features:fig:linelaserexplanation:datanoncorr} shows a series of measured laser scanner profiles of the granite table, whilst moving the profile laser scanner stepwise along the gantry in $x_{\text{isel}}$ direction. 
The data in the figure 
indicates a tilt of the profile laser scanner within the global $x,z$ plane. This tilt can be described by a simple linear relation with varying slope when moving along $x_\text{isel}$, whilst the slope is constant when moving along $y_\text{isel}$ direction. A correction was derived by measuring and averaging over multiple measurements along $y_\text{isel}$ for constant $x_\text{isel}$. 
\Figref{sec:setup:subsec:gantry:features:fig:linelaserexplanation:datacorr} shows the data in  \figref{sec:setup:subsec:gantry:features:fig:linelaserexplanation:datanoncorr} after correction. The tilt correction is applied to all measurements with the profile laser scanner.\\
{\indent}Spikes and jitters visible in Figures \ref{sec:setup:subsec:gantry:features:fig:linelaserexplanation:datanoncorr} and \ref{sec:setup:subsec:gantry:features:fig:linelaserexplanation:datacorr} are due to the changing reflectiveness of the granite table when the laser passes over grains of different colours. The height changes between the six profiles within each of the two figures are due to the gantry bridge sagging.

\section{Measurements of Wire Tension and Sag}
\label{sec:sagging:test-wires}

\subsection{Requirements on Electrode Sagging Measurements}
\label{sec:sagging:subsec:requirements:sagging}

In addition to the drift field, with its impact on NR vs ER discrimination (\secrefbra{sec:introducion}), there is also the extraction field between gate electrode and the LXe-GXe interface. The field is created by the potential difference between gate electrode and anode. To detect low-energy processes, \textit{e.g.} a low-energy nuclear recoil induced by a WIMP or CE$\nu$NS by solar neutrinos, it is crucial to operate the TPC with a high extraction efficiency ($\epsilon_\text{ext}$). 
Current experiments have achieved $\epsilon_\text{ext}$ values from $\sim\!\!\SI{50}{\%}$ to $\sim\!\!\SI{80}{\%}$ \cite{Kopec_2024}. 
An extraction field $E_{\text{ext}}$ in excess of $\SI{6}{\kilo\volt\per\centi\meter}$ in liquid xenon yields \SI{100}{\%} extraction efficiency \cite{PhysRevD.99.103024}. A potential difference between gate and anode of $\SI{7}{\kilo\volt}$ is required to achieve this 
for \SI{8}{\milli\meter} anode-to-gate-distance, a liquid level location at \SI{4}{\milli\meter} above the gate, and taking the dielectric constant of LXe as 1.96 \cite{Schmidt:2001hh}. The corresponding electric field in the xenon vapour is $\SI{11.6}{\kilo\volt\per\centi\meter}$. 
In order to achieve the desired electric field, the electrode deflection due to the electrostatic attraction between gate and anode, as well as due to the buoyancy reduced gravitational sag have to be well understood and stay in pre-defined limits. 
The electrostatic attraction between gate and anode results in a reduction of the inter-electrode distance and an increase of the electric field strength. The wire tensioning forces have to be sufficient to avoid large deflections, thereby avoiding feedback which further reduces the distance and leads to unwanted field non-uniformity. 
Starting from the forces acting on a line-element of a wire one can derive a formula estimating the wire sagging (without buoyancy):
\begin{align}
  z_{\text{min}} = z_{\text{min},\text{G}} + z_{\text{min},\text{E}} = -\frac{l_\text{wire}^2}{8\cdot T_0}\left(f_{\text{G}} + f_{\text{E}}\left(U_{\text{wire}}\right)\right)
  \label{eq:parabola:solution:diff:2dim:maxsagg}\\
    f_{\text{G}} = \rho_\text{wire} \cdot \pi \cdot r_{\text{wire}}^2 \cdot g \quad \text{and} \quad f_{\text{E}}\left(U_{\text{wire}}\right) = \lambda_\text{q}\left(U_{\text{wire}}\right) \cdot E\left(U_{\text{wire}}\right)
  \label{eq:c:fg:and:fe:all:components:general}
\end{align}
Here $f_{\text{G}}$ and $f_{\text{E}}$ are the gravitational and electrostatic force per unit length of wire. Furthermore, $r_\text{wire}$, $l_\text{wire}$, and $\rho_\text{wire}$ are the wire radius, length, and density, $g$ is the gravitational constant, $T_0$ the tensioning force, and $\lambda_\text{q}$ the charge density per unit length. For a wire with potential $U_{\text{wire}}$ suspended in parallel above a plane at zero potential in distance $d$, one can find:
\begin{align}
  f_{\text{E}} &= \frac{2 \cdot \pi \cdot \epsilon_0 \cdot \epsilon_r \cdot U_{\text{wire}}^2}{d\cdot\left(\text{arccosh}\left(\frac{d}{r_{\text{wire}}}\right)\right)^2} \label{eq:c:fe:all:components} 
\end{align}
This expression does not correspond to the electric field configuration in a TPC, but is the case for the setup in this study (\secrefbra{sec:sagging:test-wires:subsec:sagging-measurement}). These equations are valid for the limit of negligible elastic deformation due to the wire deflection and the case where the increase of $f_{\text{E}}$ as a result of the electrostatic deflection is small (\textit{i.e.} no feedback) \cite{Rolandi:2008ujz,SUDOU1996391,BALDINI2023167534}. 
We can use \eqnref{eq:parabola:solution:diff:2dim:maxsagg} to 
estimate the sagging of an anode wire of a TPC with similar size to XENONnT or LZ by choosing $l_\text{wire}=\SI{1.5}{\meter}$, $r_\text{wire}=\SI{100}{\micro\meter}$, $\rho_\text{wire}=\SI{8030}{\kilo\gram\per\meter\cubed}$ (stainless steel), $d=\SI{4}{\milli\meter}$, $T_0=\SI{10}{\newton}$ and $U_{\text{wire}}=\SI{5}{\kilo\volt}$. The maximal sag at the centre of the wire is then \SI{670}{\micro\meter} ($z_{\text{min},\text{G}}=\SI{140}{\micro\meter}$, $z_{\text{min},\text{E}}=\SI{530}{\micro\meter}$). 
In case of XENONnT, $E_\text{ext}$ takes values between \SI{2.9}{\kilo\volt} and \SI{3.7}{\kilo\volt} \cite{XENON:2024wpa}, and the corresponding distance variation is in excess of \SI{0.5}{\milli\meter}. A setup to assay electrodes for future dual-phase TPCs should therefore provide a precision better than \SI{100}{\micro\meter} to cover the relevant sagging range. Tests with such a setup could thus green-light an electrode for the next series of tests, \textit{e.g.} in a LXe environment.

\subsection{Wire Stretching and Tension Measurement}

\begin{table}
  \centering
  \begin{tabular}{l|c|c}
    quantity                     & value                                 & uncertainty                         \\ \hline \hline
    $l_{\text{wire}}$ (length)   & \SI{880}{\milli\meter}                & \SI{1}{\milli\meter}                \\
    $d$ (distance: wire-plane)   & $30.5-\SI{31.75}{\milli\meter}$       & $\sim\!\!\SI{0.6}{\milli\meter}$    \\
    $\rho_\text{wire}$ (density) & \SI{7970}{\kilo\gram\per\meter\cubed} & \SI{30}{\kilo\gram\per\meter\cubed} \\
    $r_{\text{wire}}$ (radius)   & $\SI{108}{\micro\meter}$    & \SI{1}{\micro\meter}                \\
  \end{tabular}
  \caption{\label{sec:sagging:tab:wire:props}Properties of the wires used for the sagging measurements and their uncertainties. The density value ($\rho_\text{wire}$) was obtained from the suppliers web page \cite{cfw:stainlesssteel} for stainless steel 316. As no uncertainty was given we take half the difference between the different types of stainless steel 316 offered by that supplier. The distances $d$ and their uncertainty are only approximates for all wires -- the exact value is directly measured for each wire.}
\end{table}
\begin{figure}[t]
  \centering
  \subfloat[]{\label{sec:sagging:test-wires:fig:stretched_wires:grid}\includegraphics[trim = 0 0 0 0,clip=true,width=0.45\columnwidth]{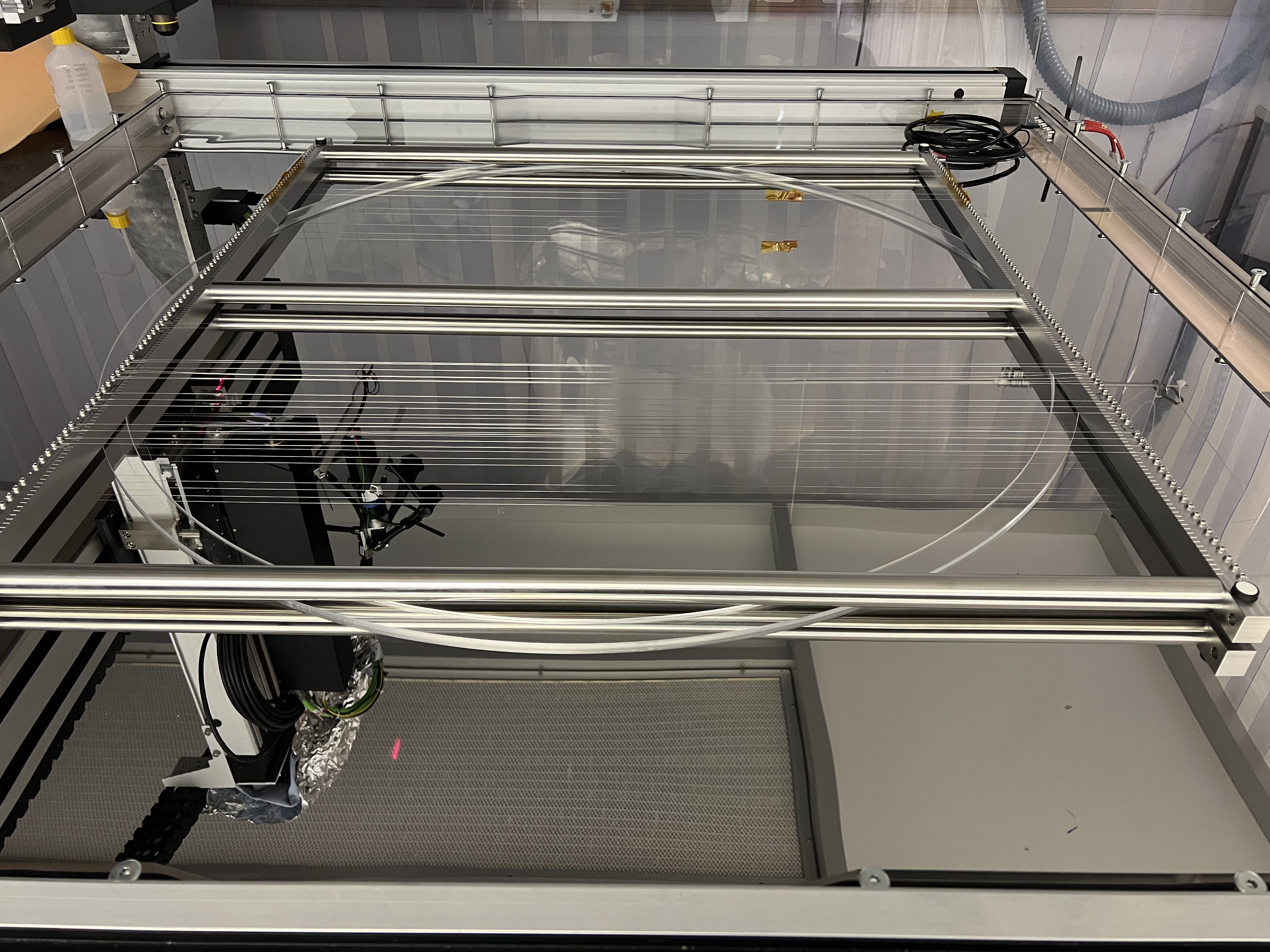}}
  \subfloat[]{\label{sec:sagging:test-wires:fig:stretched_wires:fixation}\includegraphics[trim = 0 0 0 0,clip=true,width=0.45\columnwidth]{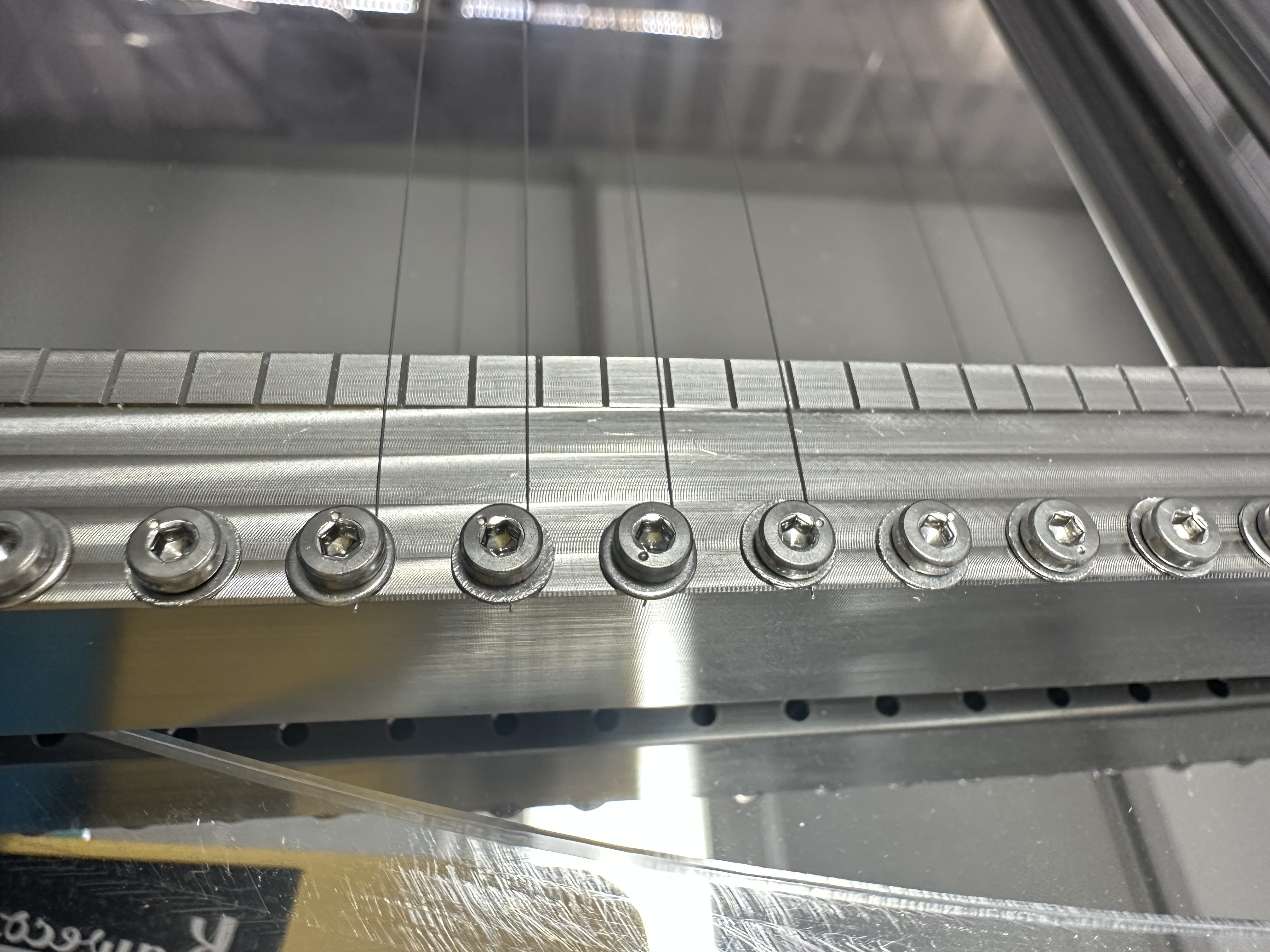}}
  \caption{\label{sec:sagging:test-wires:fig:stretched_wires}\figsubref{sec:sagging:test-wires:fig:stretched_wires:grid} The frame on which wires for the sagging measurement are stretched and fixed; the distance over which the wires hang free is \SI{880}{\milli\meter}. \figsubref{sec:sagging:test-wires:fig:stretched_wires:fixation} Detailed image of how wires are fixed with screws, before being passivated with Kapton tape. (The images do not show specific wires used during the measurement, as wires were changed during the course of the work presented here.)}
\end{figure}
Parallel wires in an electric field ($\mathcal{O}\left(\SI{1}{\kilo\volt\per\centi\meter}\right)$) are used to qualify the setup's ability to measure electrode sagging, as the small deflection due to electrostatic attraction enables the qualification of the setup at the edge of its sensitivity. The expected sagging for a simple wire geometry can be calculated straightforwardly, providing a clear validation for the measurement.
Four wires were stretched on a rectangular frame aiming at tensioning forces ($T_0$) of \SI{2}{\newton}, \SI{5}{\newton}, \SI{8}{\newton}, and \SI{10}{\newton}. The frame has a height of \SI{25}{\milli\meter}, measured to where the wires are ultimately attached, and is placed on a circular acrylic glass spacer of \SI{5}{\milli\meter} height. \Figref{sec:sagging:test-wires:fig:stretched_wires:grid} shows the frame in the test box with the polished stainless steel plate, whilst the wires' fixation is displayed in \figref{sec:sagging:test-wires:fig:stretched_wires:fixation}. Each wire was anchored with a screw on one side of the frame and then stretched using a tension spring balance, keeping the direction of the tension close to the wire's final location. Then, the other end of the wire was fixed with a second screw. The distance $d$ between ground plane and wires is measured with the profile laser scanner, albeit at a different $z_\text{isel}$ than used for the sagging measurement. Ranges for $d$ are quoted in \tabref{sec:sagging:tab:wire:props}. Given the magnitude of $d$, the relative uncertainty $(\delta d)/d$ stays small.\\
\begin{figure}
  \centering
  \includegraphics[trim = 0 0 0 0,clip=true,width=0.98\columnwidth]{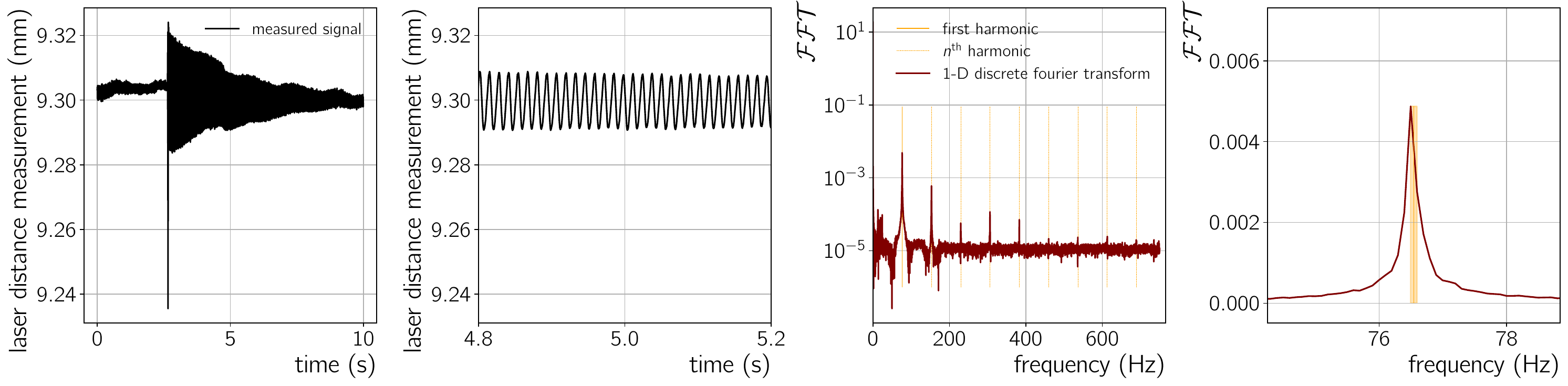}
  \caption{
  \label{sec:sagging:test-wires:fig:wire_tension}Illustration of the wire tension measurement. 
  From left to right: Distance measurement by the \textit{optoNCDT 2300-20} laser distance sensor, zoom into the first panel highlighting the oscillation of the wire, FFT of the data in the first panel where the harmonic frequencies found by our algorithm are marked, zoom into the second panel focusing at the fundamental frequency.} 
\end{figure}
{\indent}The target tensions are listed in \tabref{sec:sagging:test-wires:tab:tension-and-fh-results}. The wires were aligned parallel to the $y_{\text{isel}}$ direction of the \iselSystem{}, achieving a maximal $x_{\text{isel}}$ offset between \SI{0.5}{\milli\meter} and \SI{1.5}{\milli\meter} over their full length. 
To determine the wire tension, we excited oscillations in the wires by striking a tuning fork (\SI{100}{\hertz}) and placing it onto a corner of the frame.
Following a common technique (\textit{cf.} \cite{SUDOU1996391,BALDINI2023167534}), we calculate the wire tension ($T_0$) from the measured fundamental frequency ($f_{\text{h}}$): 
\begin{align}
  T_0 = 4 \cdot\rho_\text{wire}\cdot l_{\text{wire}}^2\cdot A_{\text{wire}} \cdot f_{\text{h}}^2
  \label{sec:sagging:test-wires:eq:wire_tension}
\end{align}
where $A_{\text{wire}}$ is the wire cross-section \cite{Rolandi:2008ujz}. The wire parameters were provided by the supplier \cite{cfw:stainlesssteel} (\textit{cf}. \tabrefbra{sec:sagging:tab:wire:props}). Their quoted accuracy on ``dimensional'' quantities is slightly worse than \SI{1}{\micro\meter}, thus we set $\delta r_{\text{wire}}=\SI{2}{\micro\meter}$. The same 
kind of wire has been used in \cite{Elykov:2025nri}. 
To measure $f_{\text{h}}$, the \textit{optoNCDT 2300-20} laser distance sensor was placed close to the centre of each wire and sampled the change in distance between laser and wire at \SI{1.5}{\kilo\hertz}. The first and second panels of \figref{sec:sagging:test-wires:fig:wire_tension} show such a \SI{10}{\second} long oscillation measurement and a \SI{0.4}{\second} zoom. Several measurements were performed per wire. The total measurement duration is about \SI{30}{\second} per wire, including moving the sensor to a known location on the wire, the manual excitation, and the measurement itself. 
The \textsc{scipy} \cite{scipy} implementation of a discrete one-dimensional fast Fourier transform (FFT) is applied to the data. Then a custom algorithm identifies $f_{\text{h}}$ as well as the subsequent harmonics (\figrefbra{sec:sagging:test-wires:fig:wire_tension}, third and fourth panel). The final result for $f_{\text{h}}$ is the mean of all identified harmonics, divided by their order number (\textit{e.g.} divided by 2 for the second harmonic). The uncertainty $\delta f_{\text{h}}$ is taken as the standard deviation divided by the number of harmonics identified in the measurement. 
As a last step, the weighted mean \cite{stats_weighted} ($\overline{f_{\text{h}}}$) of the ten $f_{\text{h}},\delta f_{\text{h}}$ value pairs per wire and its uncertainty ($\delta \overline{f_{\text{h}}}$) are calculated using $(\delta f_{\text{h}}^i)^{-2}$ as weight.\\
\begin{table}
  \centering
  \begin{tabular}{c|c|c|c|c|c}
    wire nb. & target $T_0$   & $\overline{f_{\text{h}}}\pm\delta \overline{f_{\text{h}}}$ & $T_0 \pm \delta T_0$  & $z_{\text{min,G}}\pm\delta z_{\text{min,G}}$ & $z_{\text{min}}\pm\delta z_{\text{min}}$ \\ 
             & (\si{\newton}) & (\si{\hertz})                                              & (\si{\newton})        & (\si{\micro\meter})                          & (\si{\micro\meter})                      \\ \hline \hline
    1 & 10 & $ 107.25 \pm 0.06$ & $ 10.4\pm0.4$ & $-27\pm1$ & $\ -42\pm\ 4$ \\
    2 &  8 & $ 108.66 \pm 0.03$ & $ 10.7\pm0.4$ & $-26\pm1$ & $\ -41\pm\ 4$ \\
    3 &  5 & $\ 79.72 \pm 0.02$ & $\ 5.8\pm0.2$ & $-48\pm3$ & $\ -76\pm\ 7$ \\
    4 &  2 & $\ 61.22 \pm 0.03$ & $\ 3.4\pm0.1$ & $-82\pm4$ & $ -129\pm 11$ \\
  \end{tabular}
  \caption{\label{sec:sagging:test-wires:tab:tension-and-fh-results}Weighted mean ($\overline{f_{\text{h}}}\pm\delta \overline{f_{\text{h}}}$) of the measured fundamental frequencies of all wires as well as the wire tension ($T_0 \pm \delta T_0$) calculated from \eqnref{sec:sagging:test-wires:eq:wire_tension} using the frequencies in this table and the values in \tabref{sec:sagging:tab:wire:props}. The uncertainty $\delta T_0$ is dominated by the uncertainty on the wire properties. $z_{\text{min,G}}\pm \delta z_{\text{min,G}}$ and $z_{\text{min}}\pm \delta z_{\text{min}}$ are the expected sagging amplitude due to gravity and the total sagging (including gravitation and electric fields for $U_\text{wire}=\SI{6000}{\volt}$), respectively. These values are calculated using Equations \eqref{eq:parabola:solution:diff:2dim:maxsagg}, \eqref{eq:c:fg:and:fe:all:components:general} and \eqref{eq:c:fe:all:components}, and the information in \tabref{sec:sagging:tab:wire:props}.}
\end{table}
{\indent}The wire tension $T_0$ is determined by using $\overline{f_{\text{h}}}\pm\delta \overline{f_{\text{h}}}$, the wire properties listed in \tabref{sec:sagging:tab:wire:props}, and \eqnref{sec:sagging:test-wires:eq:wire_tension}. Gaussian error propagation is used to calculate the uncertainty $\delta T_0$. \Tabref{sec:sagging:test-wires:tab:tension-and-fh-results} summarises the weighted mean of the measured frequencies and the calculated tension for every wire. The measured $T_0$ scatter around the desired tension, \textit{cf}. \textit{target $T_0$} in \tabref{sec:sagging:test-wires:tab:tension-and-fh-results}. During initial tests, the experience was made that some tension is lost when fixing the wires with the screws. Compensating for this effect resulted for most cases in slightly larger $T_0$. The range of tensions is sufficiently broad to quantify the performance of our setup.

\subsection{Sagging Measurement}
\label{sec:sagging:test-wires:subsec:sagging-measurement}

\begin{figure}
  \centering
  \subfloat[]{\label{sec:sagging:test-wires:subsec:sagging-measurement:fig:example:profiles:frame}\includegraphics[width=0.49\columnwidth]{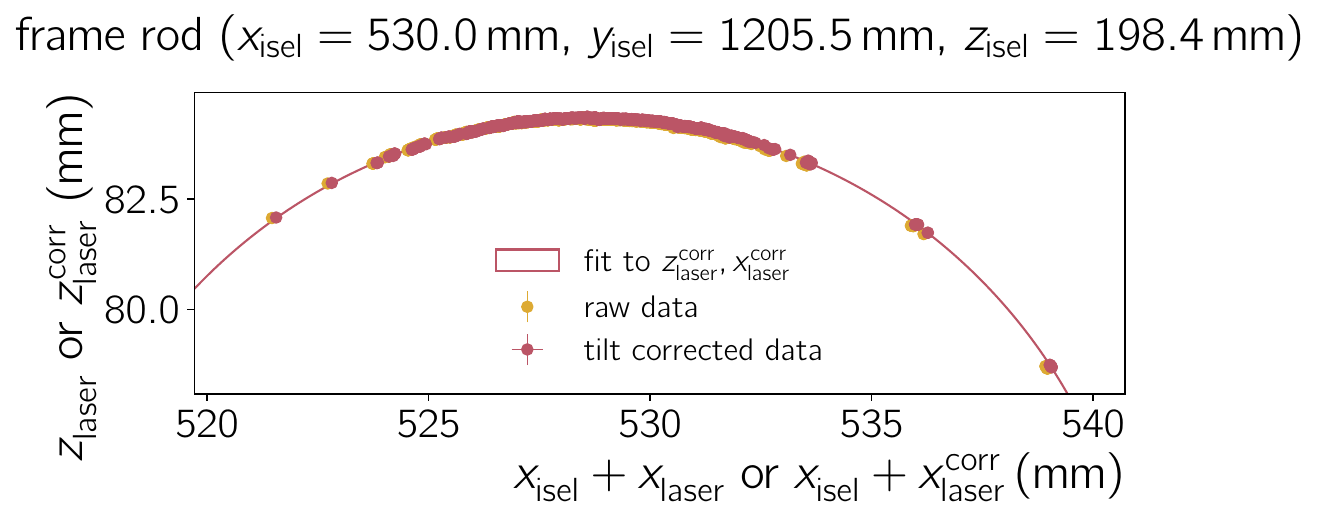}}
  \subfloat[]{\label{sec:sagging:test-wires:subsec:sagging-measurement:fig:example:profiles:wire}\includegraphics[width=0.49\columnwidth]{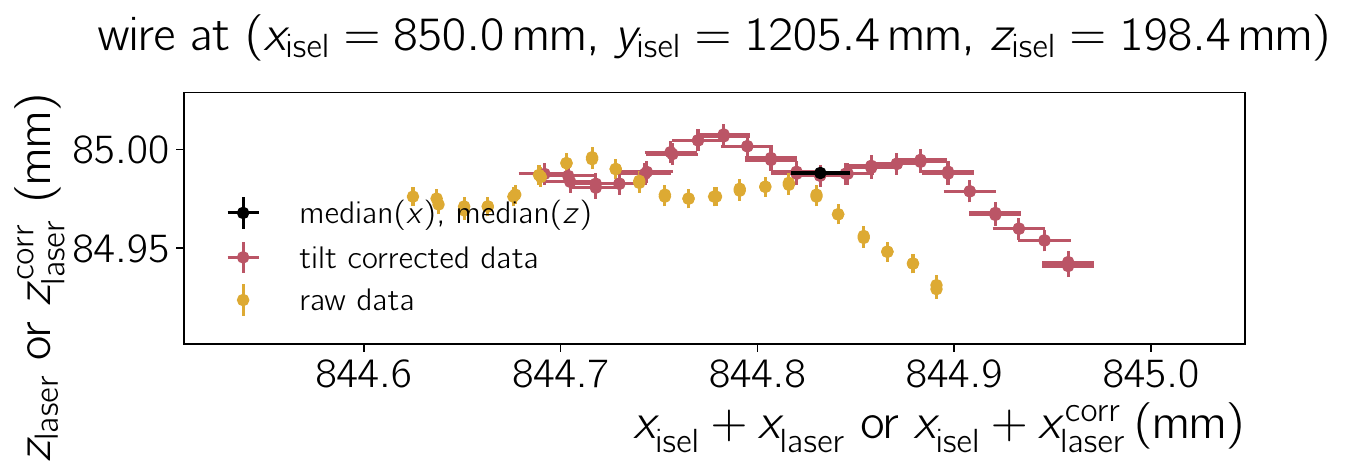}}
  \caption{\label{sec:sagging:test-wires:subsec:sagging-measurement:fig:example:profiles}A profile measured by the laser scanner of \figsubref{sec:sagging:test-wires:subsec:sagging-measurement:fig:example:profiles:frame} a rod of the frame on which the wires are mounted, and \figsubref{sec:sagging:test-wires:subsec:sagging-measurement:fig:example:profiles:wire} of one of the wires.}
\end{figure}
Sagging measurements are ``elevation profiles'' of the height of the wire vs. a coordinate parallel to the wire. For this set-up, the latter coordinate is $y_\text{isel}$, due to the almost parallel alignment of the wires with the $y_\text{isel}$ direction of better than $0.1\,^{\circ}$. The height of the wire ($z_\text{wire}$) is extracted from wire profiles measured in few-\si{\milli\meter} intervals perpendicular to the wire direction. For \SI{1}{\milli\meter} steps in between measurement positions, it takes less than 45 minutes to scan along \SI{1}{\meter} of $y_\text{isel}$. Given the $\sim\!\!\SI{2.5}{\centi\meter}$ width of the \textit{scanCONTROL 3000} laser-profile in $x_\text{isel}$, profiles of several wires can be included in the same recording. For the measurement, the corrections discussed in \secref{sec:setup:subsec:gantry:features} are applied, and \iselSystem{} movement steps are kept the same. 
The same procedure is used to perform measurements of the frame's three rods (see the horizontal bars in \figrefbra{sec:sagging:test-wires:fig:stretched_wires:grid}). Different settings of the profile laser scanner were chosen for the measurement of the wires (exposure time: \SI{2}{\milli\second}, line rate: \SI{250}{\hertz}) and the frame rods (exposure time: \SI{0.2}{\milli\second}, line rate: \SI{250}{\hertz}). A longer exposure time is needed to resolve the wires properly, whilst the shorter time chosen for the rod-profiles reduces glare effects.\\
{\indent}The rod-profiles serve as a reference measurement of the gantry induced distance offset, recorded close in time to the wire sagging measurement. This data is also used to monitor that the frame did not tip due to the electrostatic attraction between grounded plane and the frame when HV is applied. 
\Figref{sec:sagging:test-wires:subsec:sagging-measurement:fig:example:profiles:frame} shows an example profile of a frame rod, with a circle fitted to the data. Such a circle is fitted to all rod profiles. Profiles where the fit failed ($\sim\!\!\SI{17}{\%}$) are excluded. Most failed fits correspond to rod-profiles with still present glare effects, distorting the expected circular shape. 
The average rod radius ($r_\text{rod}^{\circ\text{-fit}}$) of all fits passing the cuts is \SI{12.45(3)}{\milli\meter}, using the standard deviation as uncertainty. 
The expected rod radius is \SI{12.5}{\milli\meter}.\\[0.2cm]
{\indent}The example wire profile in \figref{sec:sagging:test-wires:subsec:sagging-measurement:fig:example:profiles:wire} shows how the laser scanner struggles to measure clear wire profiles with their \SI{216}{\micro\meter} diameter. 
Profiles of wires may seem erroneously broadened as a result of the wires' curvature, especially at the edges in $x_\text{laser}$ direction \cite{Line_Laser_Broadened_Error:2024}. The expected wire geometry informs conservative quality cuts on the measured profiles, where \SI{76.8}{\%} of profiles pass the cuts. 
Profiles excluded are predominantly located at the very end of the wires, where reflections from the frame distort the measurement. 
The width of the wire 
is defined as $\text{max}\left(x_\text{laser}^\text{corr}\right)-\text{min}\left(x_\text{laser}^\text{corr}\right)$, where the $x_\text{laser}^\text{corr}$ are all the corresponding values measured in one profile. 
The mean 
width and its standard deviation are \SI{216}{\micro\meter} and \SI{21}{\micro\meter}, respectively, when taking all profiles passing the cuts into account. This width fits the expectation but the large standard deviation is in line with what has been discussed for the exemplary wire profile in \figref{sec:sagging:test-wires:subsec:sagging-measurement:fig:example:profiles:wire}. 
For each profile passing the cuts the measured height of the wire is defined to be $z_\text{wire}=\text{median}\left(z_\text{laser}^\text{corr}\right)$ in the profile's position. The uncertainty of this height ($\delta z_\text{wire}$) is calculated as the uncertainty on the median $\delta z_\text{wire} = \sigma_{z_\text{laser}^\text{corr}} \cdot \sqrt{0.5 \cdot \pi \cdot N_\text{laser}^{-1}}$, where $\sigma_{z_\text{laser}^\text{corr}}$ is the standard deviation and $N_\text{laser}^{-1}$ the number of data points in the respective profile measured by the laser scanner.
The uncertainty of the $x$ position of the wire is defined analogously, whilst the position is defined as $x_\text{wire}=\text{median}\left(x_\text{laser}^\text{corr}\right)+x_\text{isel}$. The median and 
its uncertainty are more stable against occasional outliers than \textit{e.g.} the mean or the maximal $z_\text{laser}$ value.

\subsubsection*{Absolute Sagging Measurement}
\label{sec:sagging:test-wires:subsec:sagging-measurement:subsec:abs}

\begin{figure}
  \centering
  \subfloat[]{\label{sec:sagging:test-wires:subsec:sagging-measurement:subsec:abs:fig:explain:meas:10N:raw}\includegraphics[width=0.32\columnwidth,trim=0 0 0 32,clip=true]{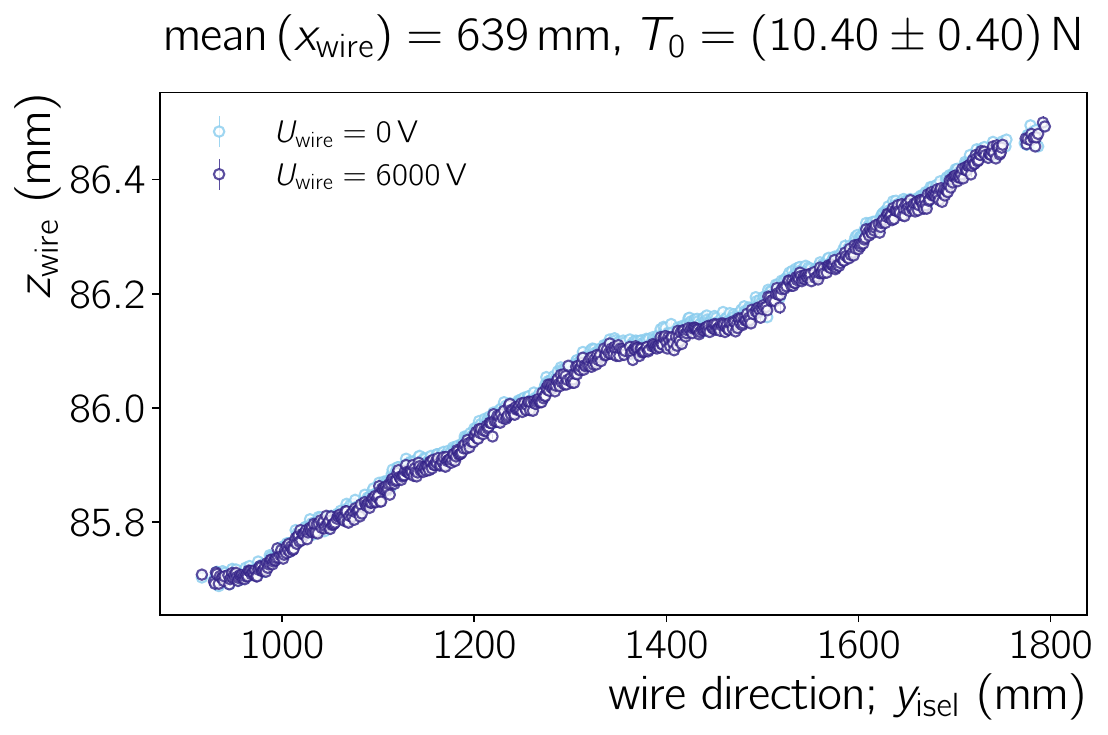}}
  \subfloat[]{\label{sec:sagging:test-wires:subsec:sagging-measurement:subsec:abs:fig:explain:meas:10N:P1corr}\includegraphics[width=0.32\columnwidth,trim=0 0 0 32,clip=true]{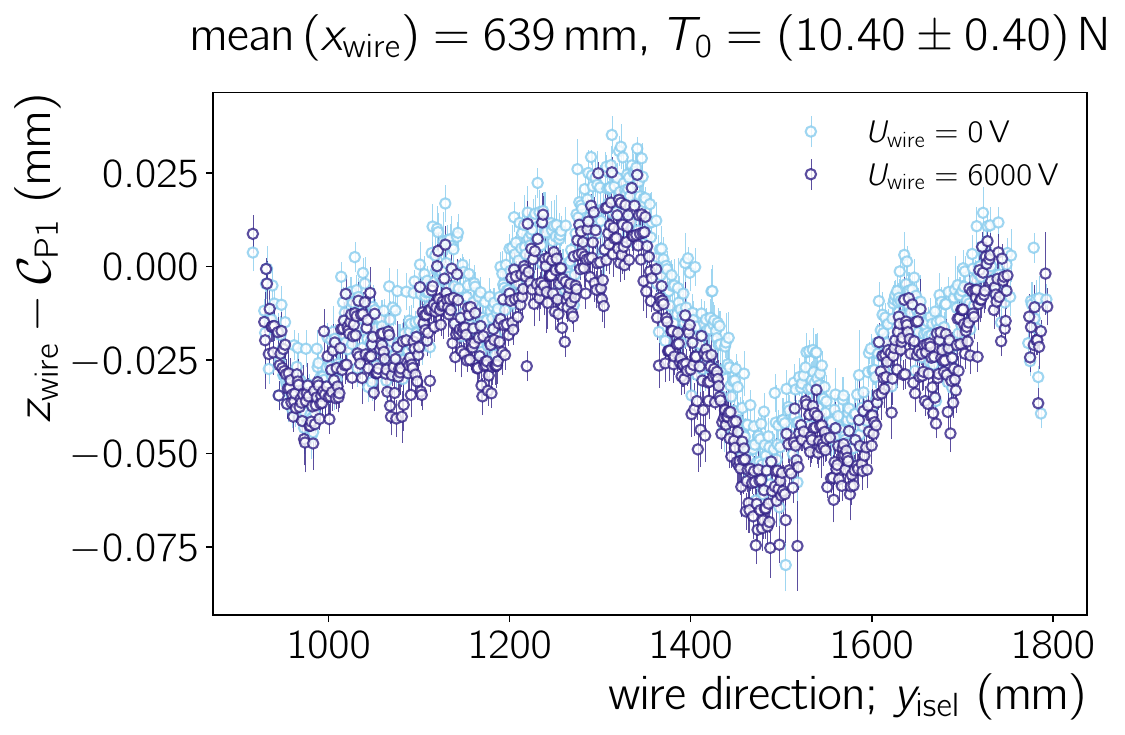}}
  \subfloat[]{\label{sec:sagging:test-wires:subsec:sagging-measurement:subsec:abs:fig:explain:meas:10N:P1corr:sincorr:grav:p10fit}\includegraphics[width=0.32\columnwidth,trim=0 0 0 32,clip=true]{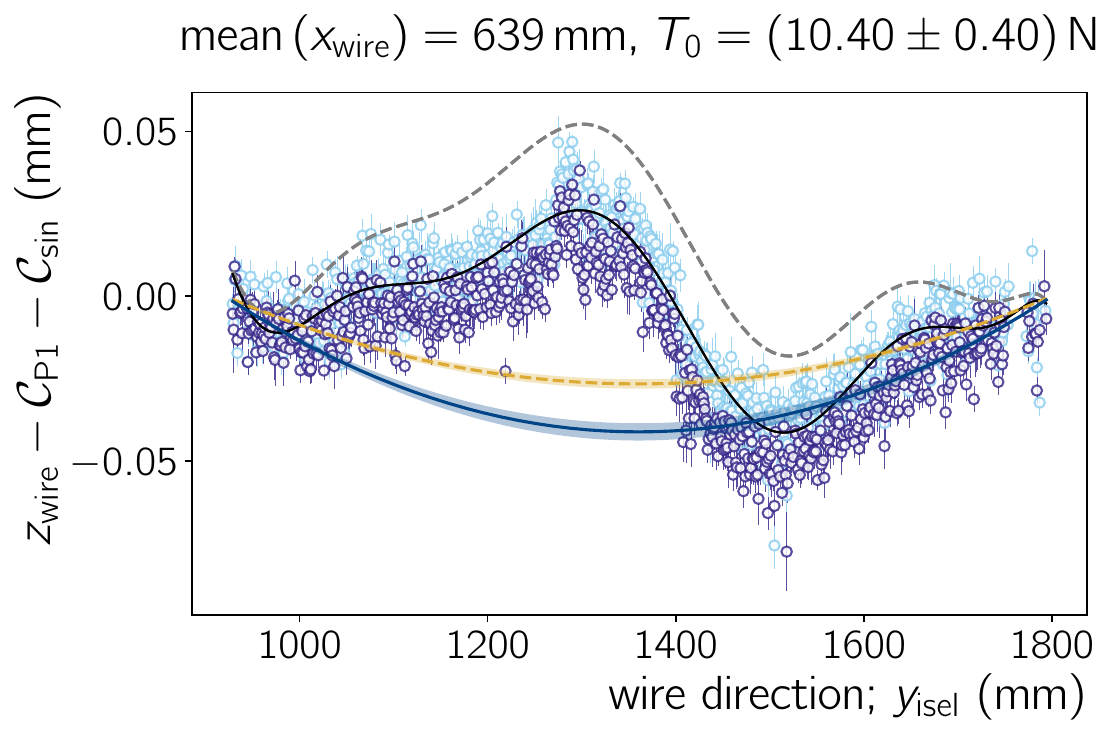}}\\
  \subfloat[]{\label{sec:sagging:test-wires:subsec:sagging-measurement:subsec:abs:fig:explain:meas:3N:raw}\includegraphics[width=0.32\columnwidth,trim=0 0 0 32,clip=true]{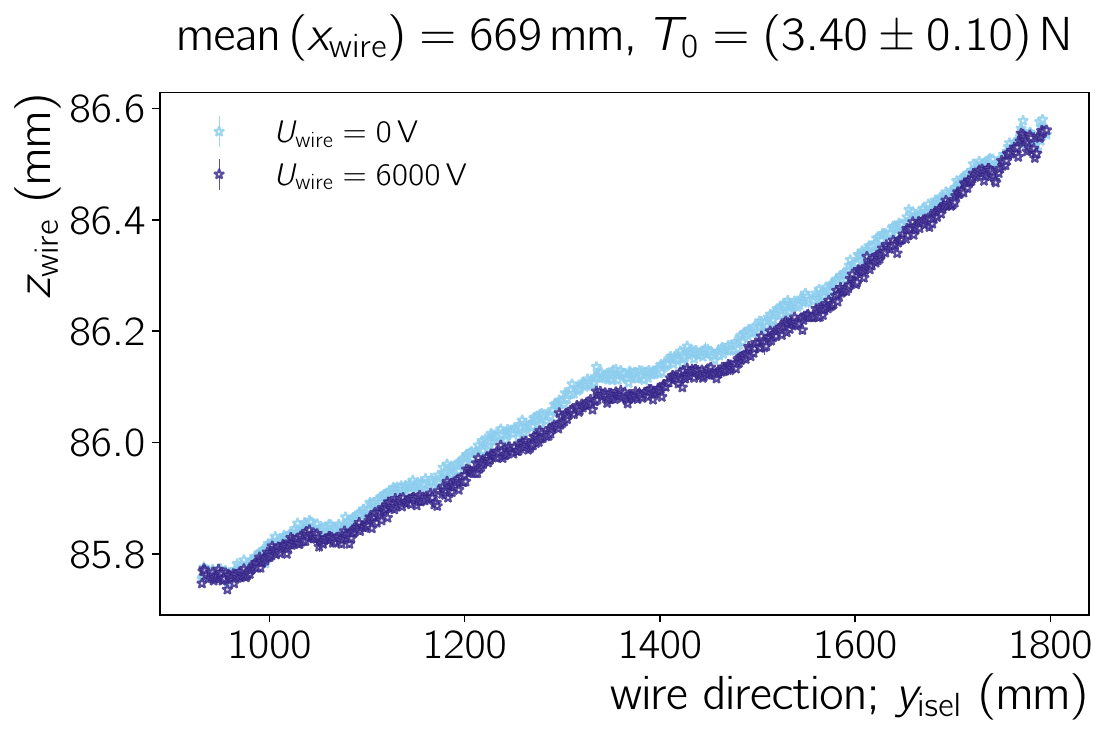}}
  \subfloat[]{\label{sec:sagging:test-wires:subsec:sagging-measurement:subsec:abs:fig:explain:meas:3N:P1corr}\includegraphics[width=0.32\columnwidth,trim=0 0 0 32,clip=true]{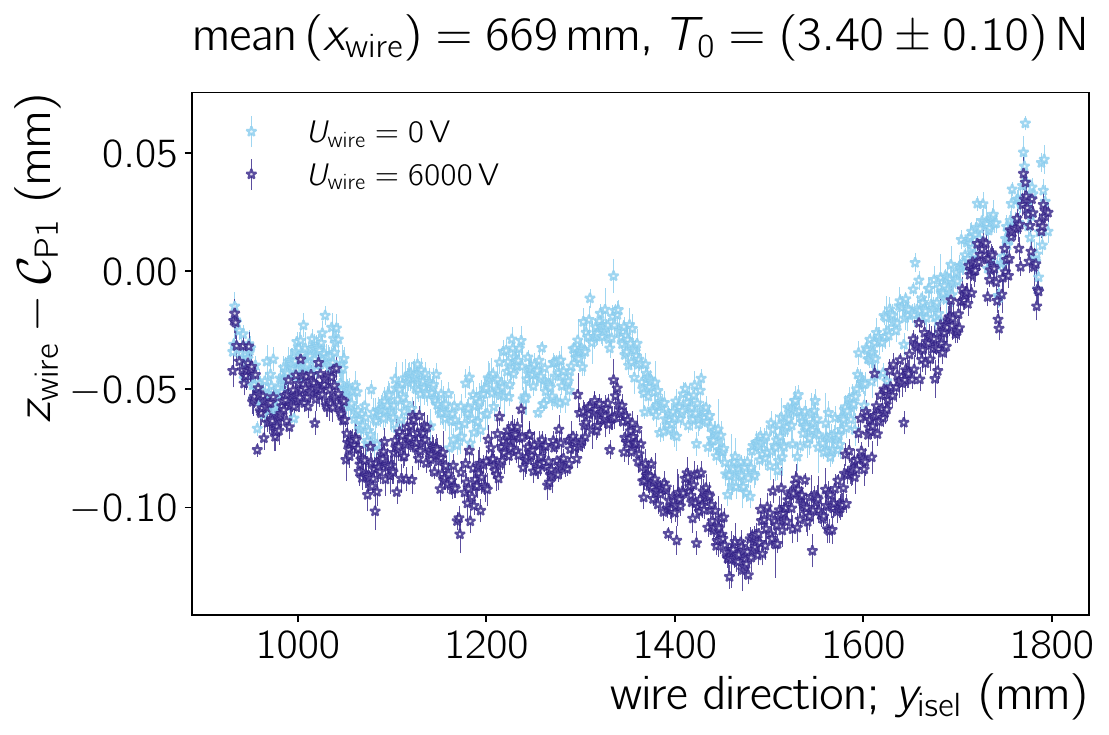}}
  \subfloat[]{\label{sec:sagging:test-wires:subsec:sagging-measurement:subsec:abs:fig:explain:meas:3N:P1corr:sincorr:grav}\includegraphics[width=0.32\columnwidth,trim=0 0 0 32,clip=true]{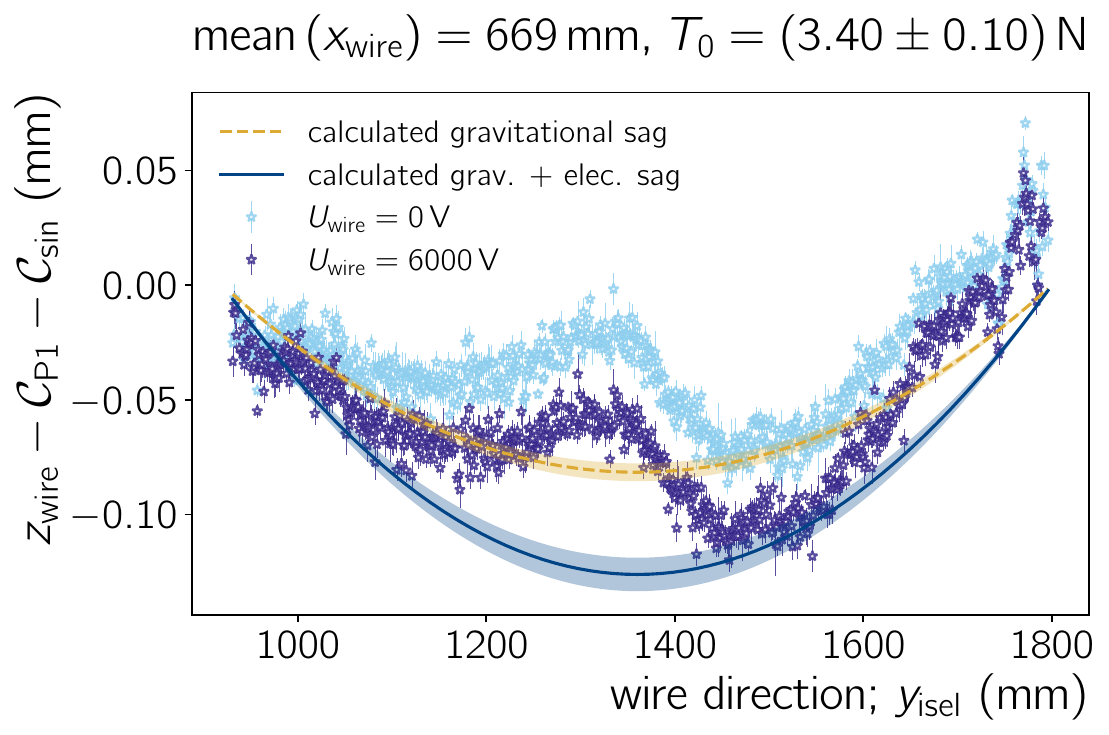}}
  \caption{\label{sec:sagging:test-wires:subsec:sagging-measurement:subsec:abs:fig:explain:meas}First row: Wire at position $\text{mean}\left(x_\text{wire}\right)=\SI{639}{\milli\meter}$, tensioned with $T_0=\SI{10.4(4)}{\newton}$. \figsubref{sec:sagging:test-wires:subsec:sagging-measurement:subsec:abs:fig:explain:meas:10N:raw} Raw elevation profile along the wire as measured, \figsubref{sec:sagging:test-wires:subsec:sagging-measurement:subsec:abs:fig:explain:meas:10N:P1corr} the same data after subtracting the tilt of the frame ($\mathcal{C}_\text{P1}$ correction), and \figsubref{sec:sagging:test-wires:subsec:sagging-measurement:subsec:abs:fig:explain:meas:10N:P1corr:sincorr:grav:p10fit} the $\mathcal{C}_\text{P1}$ corrected data after subtracting the oscillations ($\mathcal{C}_\text{sin}$ correction). The last plot also displays the expected sagging curve for only the gravitational force as well as the gravitational force and the electrostatic force combined. The black solid line is a fit to the $U_\text{wire}=\SI{0}{\volt}$ data, whilst the grey dashed line shows the same fit but subtracted with the expected gravitational sagging. Second row: Wire at position $\text{mean}\left(x_\text{wire}\right)=\SI{669}{\milli\meter}$, tensioned with $T_0=\SI{3.4(1)}{\newton}$. The plots \figsubref{sec:sagging:test-wires:subsec:sagging-measurement:subsec:abs:fig:explain:meas:3N:raw}, \figsubref{sec:sagging:test-wires:subsec:sagging-measurement:subsec:abs:fig:explain:meas:3N:P1corr}, and \figsubref{sec:sagging:test-wires:subsec:sagging-measurement:subsec:abs:fig:explain:meas:3N:P1corr:sincorr:grav} are analogues to the plots right above them.}
\end{figure}
This section details the necessary corrections of the raw elevation profiles, so that the absolute sagging can be estimated, using the wires with the highest and lowest tension. These corrections rely for the most part on the wire data itself.
Plotting the $z_\text{wire}$ data vs $y_\text{isel}$ shows the raw elevation profile of a wire (\figrefbra{sec:sagging:test-wires:subsec:sagging-measurement:subsec:abs:fig:explain:meas:10N:raw} and \figrefbra{sec:sagging:test-wires:subsec:sagging-measurement:subsec:abs:fig:explain:meas:3N:raw}).
When looking at the figures, there is first a $\sim\!\!\SI{0.8}{\milli\meter}$ height difference between the raw elevation profiles' beginning and end, because the ground plane is higher on one side -- most likely due to the glue layer fixing the plane to the floor of the acrylic box being thicker there. The resulting angle is less than $0.06\,^{\circ}$ for all wires. 
This tilt is subtracted using a first order polynomial (P1) from the $z_\text{wire}$ values 
($\mathcal{C}_\text{P1}$ correction, \figrefbra{sec:sagging:test-wires:subsec:sagging-measurement:subsec:abs:fig:explain:meas:10N:P1corr} and \figrefbra{sec:sagging:test-wires:subsec:sagging-measurement:subsec:abs:fig:explain:meas:3N:P1corr}). The parameters of the P1 have been determined by measuring the height wire frame on both ends of each wire, and then interpolating between these points.\\
{\indent}The result after $\mathcal{C}_\text{P1}$ correction reveals various features introduced by the gantry axis and the way how they are mounted to the table. A correction of the sinusoidal oscillations has been developed using elevation profiles of the wire frame's rods along $y_\text{isel}$. The rod data was taken at the same $z_\text{isel}$ position and using the same $y_\text{isel}$ steps/positions as for the measurements of the wires, albeit being located at a different $x_\text{isel}$ location. 
The tilt of the frame rod is first subtracted, similar to the $\mathcal{C}_\text{P1}$ correction of the wire elevation profiles. Next, a pre-fit of the sinusoidal component is performed on the corrected rod elevation profile. The result is used to subtract the oscillations from the data. After this step, the resulting elevation profile is fitted with a $10^\text{th}$ order polynomial (P10) using the \texttt{numpy.polynomial} package. Then the previously subtracted sinusoidal is added again and the fitted P10 is subtracted from that data. A reapplied sinusoidal fit yields oscillations which repeat every \SI{100.2(1)}{\milli\meter}, with an amplitude of \SI{9.5(1)}{\micro\meter}, and a phase offset of \SI{0.01(3)}{\milli\meter}. 
The oscillations remain the same when moving the gantry in $x_\text{isel}$ or $z_\text{isel}$, provided the same movement steps over the exact same $y_\text{isel}$ positions are performed. Thus, these parameters are used to subtract the \iselSystem{} induced oscillations from the wires' elevation profiles ($\mathcal{C}_\text{sin}$ correction).\\
{\indent}The elevation profiles after $\mathcal{C}_\text{P1}$ and $\mathcal{C}_\text{sin}$ subtraction are shown in \figref{sec:sagging:test-wires:subsec:sagging-measurement:subsec:abs:fig:explain:meas:10N:P1corr:sincorr:grav:p10fit} and \figref{sec:sagging:test-wires:subsec:sagging-measurement:subsec:abs:fig:explain:meas:3N:P1corr:sincorr:grav} for the $T_0=\SI{10.4(4)}{\newton}$ and $T_0=\SI{3.4(1)}{\newton}$ wire, respectively. The difference between the $U_\text{wire}=\SI{6000}{\volt}$ data points and $U_\text{wire}=\SI{0}{\volt}$ data points is clearly visible. Overlaid is a calculation of the expected sagging for the gravitational force only ($f_{\text{E}}\left(U_\text{wire}\right)=0$) as well as for the gravitational and electrostatic force:
\begin{align}
  z_\text{wire}\left(y_\text{isel},U_\text{wire}\right) &= \frac{1}{2}\cdot\frac{f_{\text{G}}+f_{\text{E}}\left(U_\text{wire}\right)}{T_0}\cdot\left(\left(y_\text{isel}-y_\text{isel}^\text{shift}\right)^2 - \frac{l^2}{4}\right)
  \label{sec:sagging:test-wires:subsec:sagging-measurement:eq:sagging:expectation}
\end{align}
\begin{figure}
  \centering
  \subfloat[]{\label{sec:sagging:test-wires:subsec:sagging-measurement:subsec:abs:fig:explain:meas:10N:P1corr:sincorr:p10corr}\includegraphics[width=0.32\columnwidth]{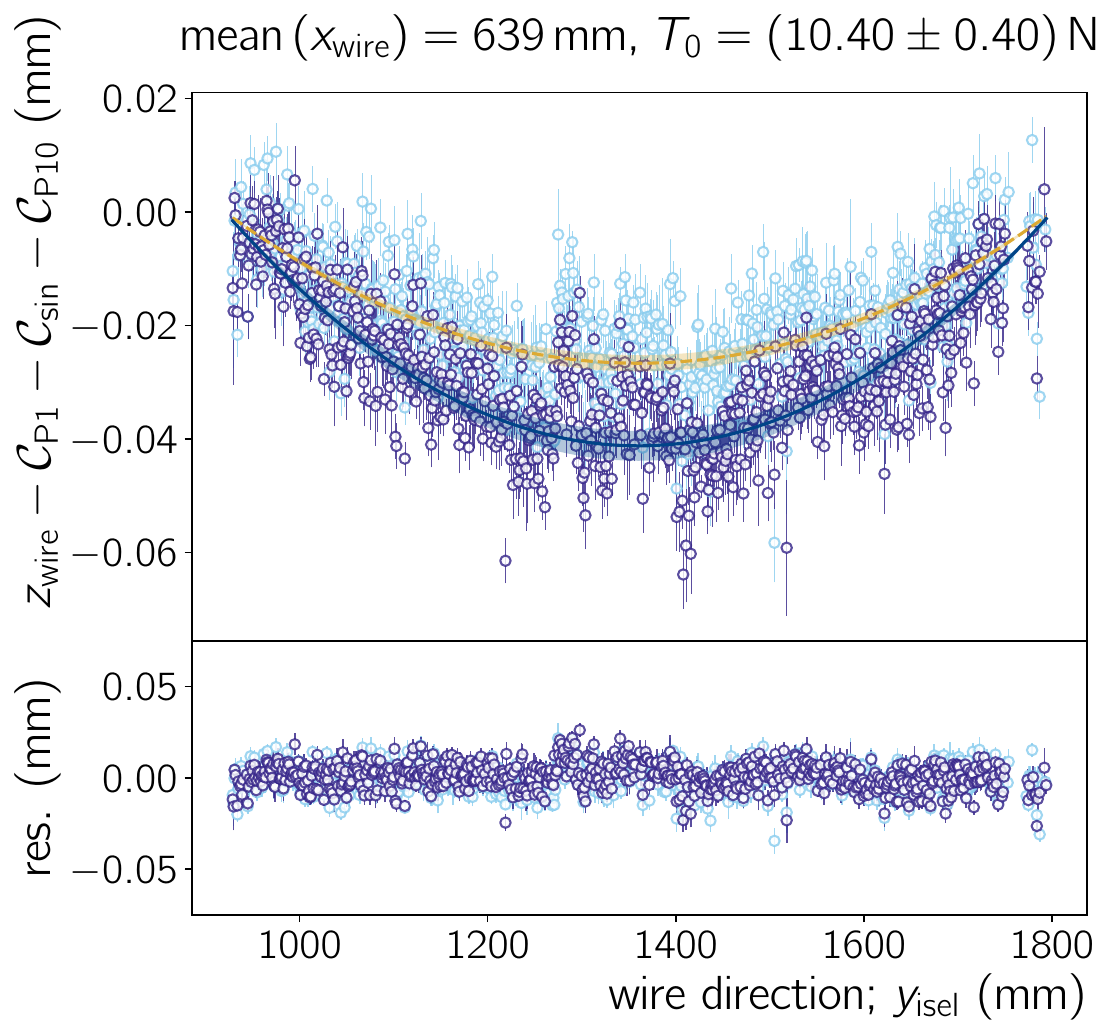}}\hspace{0.2cm}
  \subfloat[]{\label{sec:sagging:test-wires:subsec:sagging-measurement:subsec:abs:fig:explain:meas:10N:2:P1corr:sincorr:p10corr}\includegraphics[width=0.32\columnwidth]{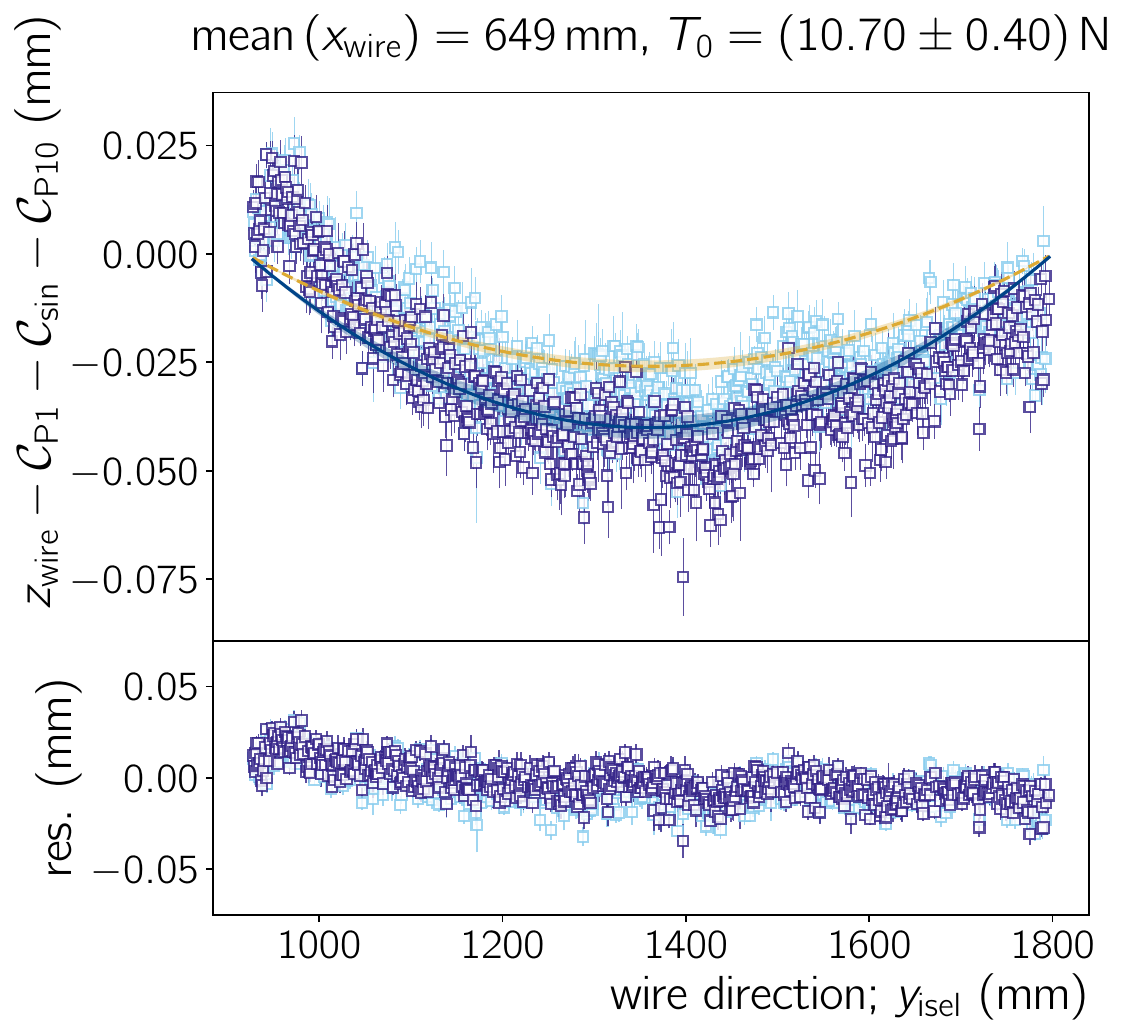}}\\[0.2cm]
  \subfloat[]{\label{sec:sagging:test-wires:subsec:sagging-measurement:subsec:abs:fig:explain:meas:5N:P1corr:sincorr:p10corr}\includegraphics[width=0.32\columnwidth]{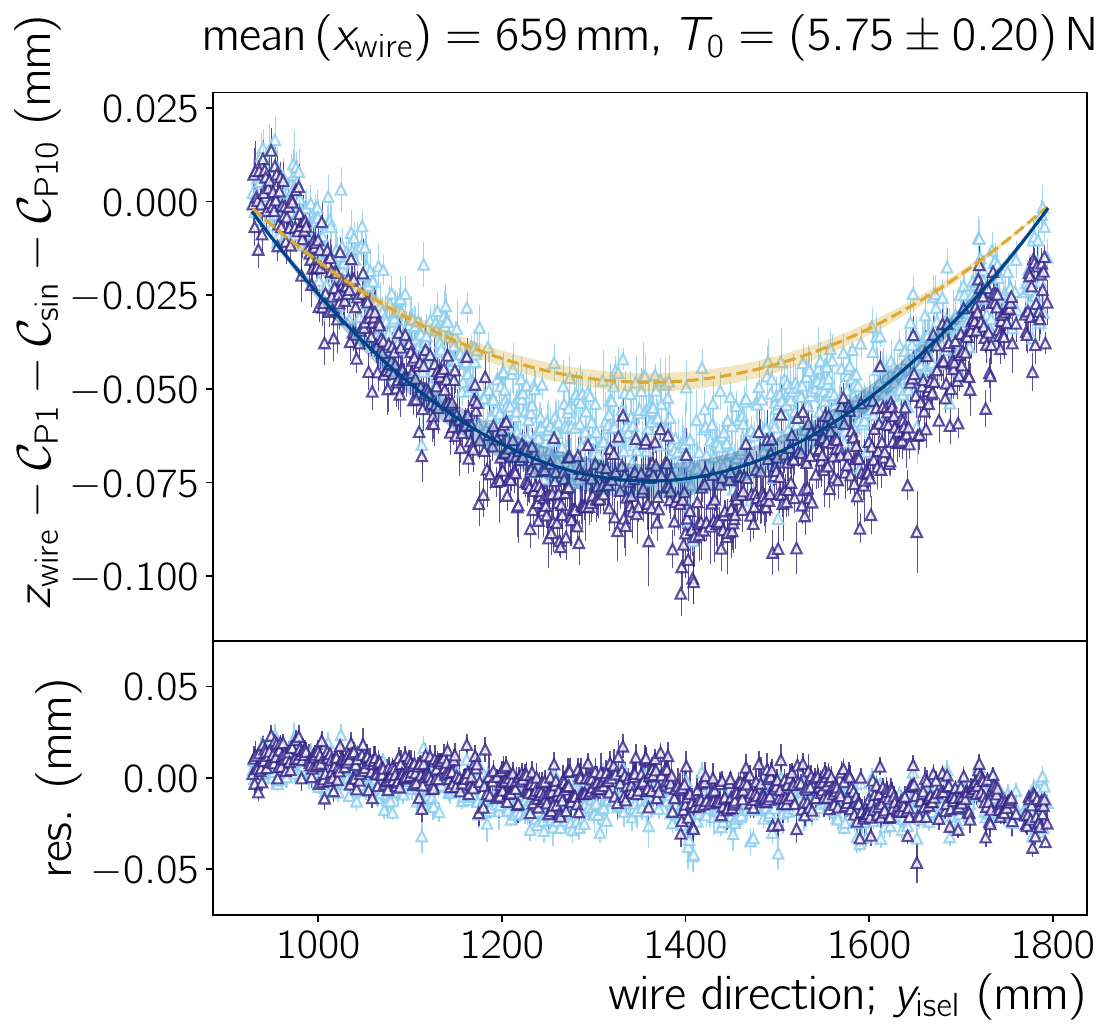}}\hspace{0.2cm}
  \subfloat[]{\label{sec:sagging:test-wires:subsec:sagging-measurement:subsec:abs:fig:explain:meas:3N:P1corr:sincorr:p10corr}\includegraphics[width=0.32\columnwidth]{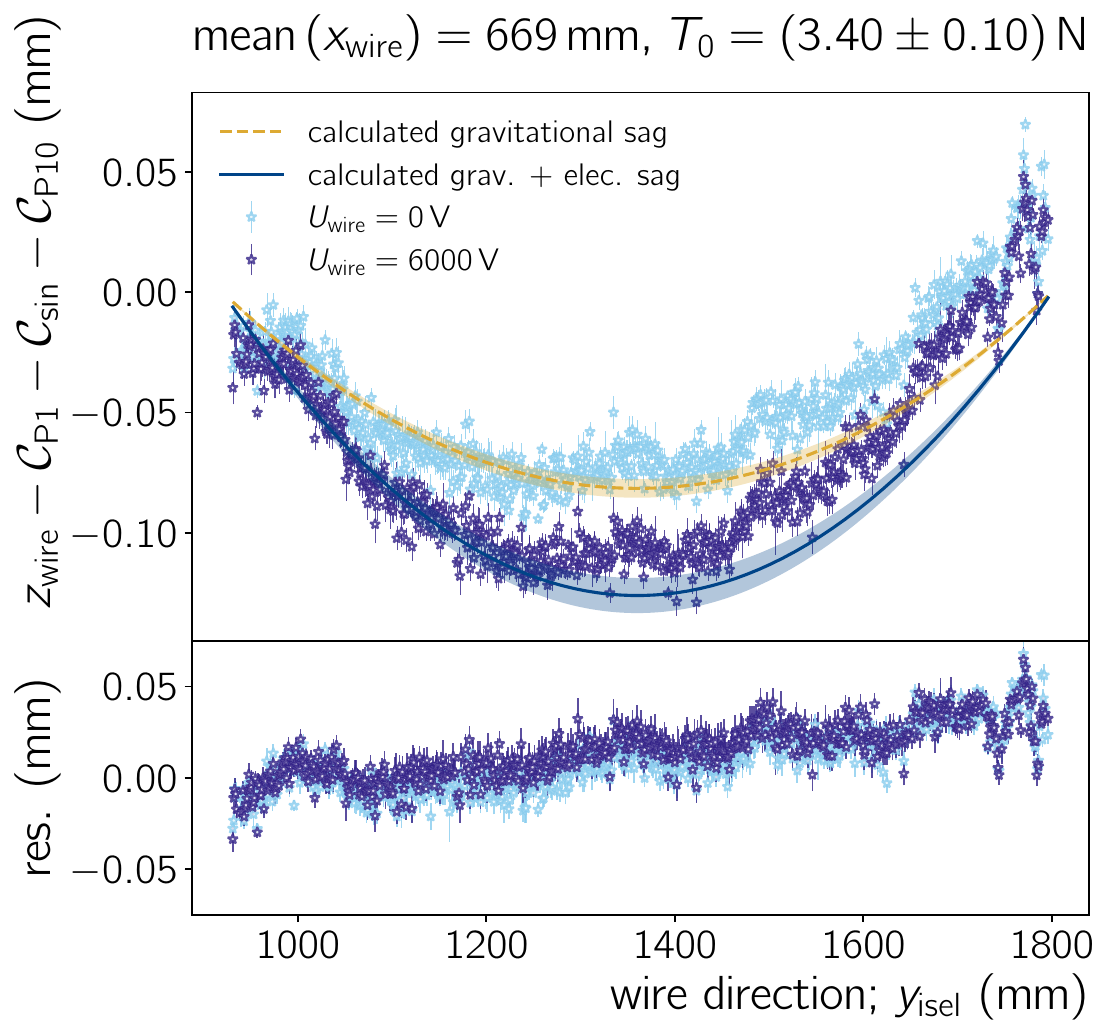}}
  \caption{\label{sec:sagging:test-wires:subsec:sagging-measurement:subsec:abs:fig:explain:meas:P1corr:sincorr:p10corr}Corrected elevation profiles of four different wires with varying tension, as indicated above each plot. The legend in \figsubref{sec:sagging:test-wires:subsec:sagging-measurement:subsec:abs:fig:explain:meas:3N:P1corr:sincorr:p10corr} applies to all plots. The corrections ($\mathcal{C}_\text{P1}$, $\mathcal{C}_\text{sin}$ and $\mathcal{C}_\text{P10}$) are explained in the text. The wires in \figsubref{sec:sagging:test-wires:subsec:sagging-measurement:subsec:abs:fig:explain:meas:10N:P1corr:sincorr:p10corr} and \figsubref{sec:sagging:test-wires:subsec:sagging-measurement:subsec:abs:fig:explain:meas:3N:P1corr:sincorr:p10corr} are the same ones as displayed in the first and second row, respectively, of \figref{sec:sagging:test-wires:subsec:sagging-measurement:subsec:abs:fig:explain:meas}. The lower panel in every plot shows the residuals between the $U_\text{wire}=\SI{0}{\volt}$ (\SI{6000}{\volt}) data and the calculated gravitational (gravitational and electrostatic) sagging.}
\end{figure}
\begin{figure}
  \centering
  \subfloat[]{\label{sec:sagging:test-wires:subsec:sagging-measurement:subsec:electro:fig:meas:10N}\includegraphics[width=0.375\columnwidth]{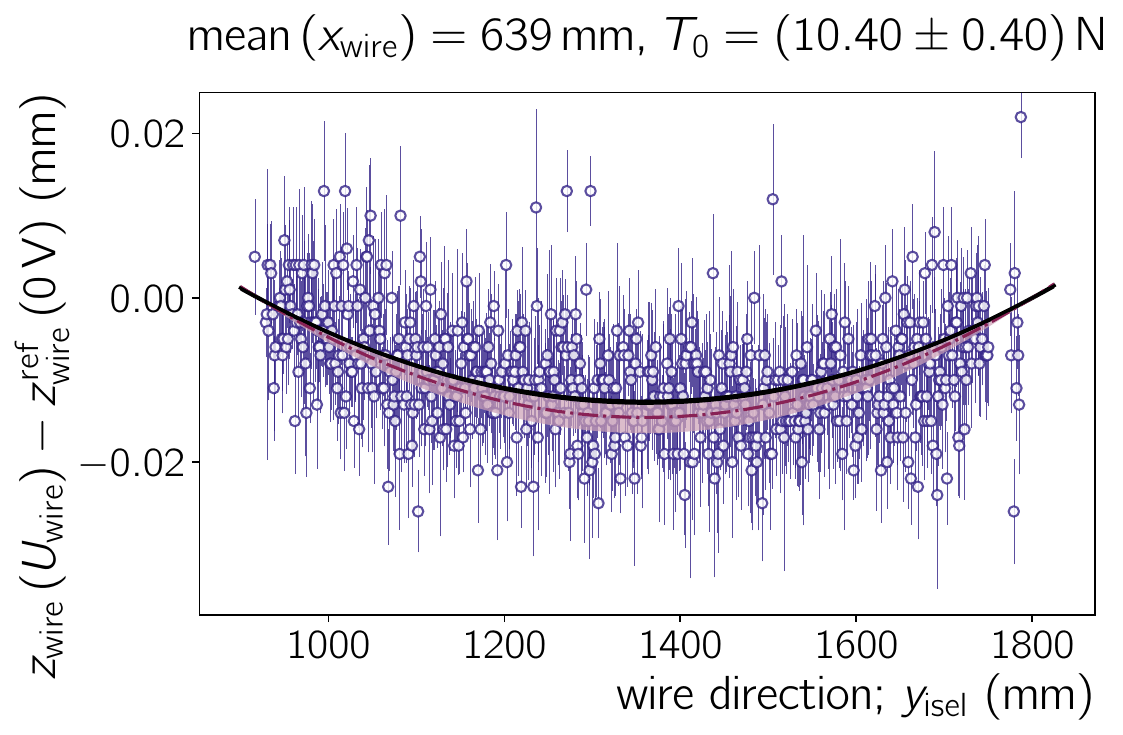}}
  \subfloat[]{\label{sec:sagging:test-wires:subsec:sagging-measurement:subsec:electro:fig:meas:10N:2}\includegraphics[width=0.375\columnwidth]{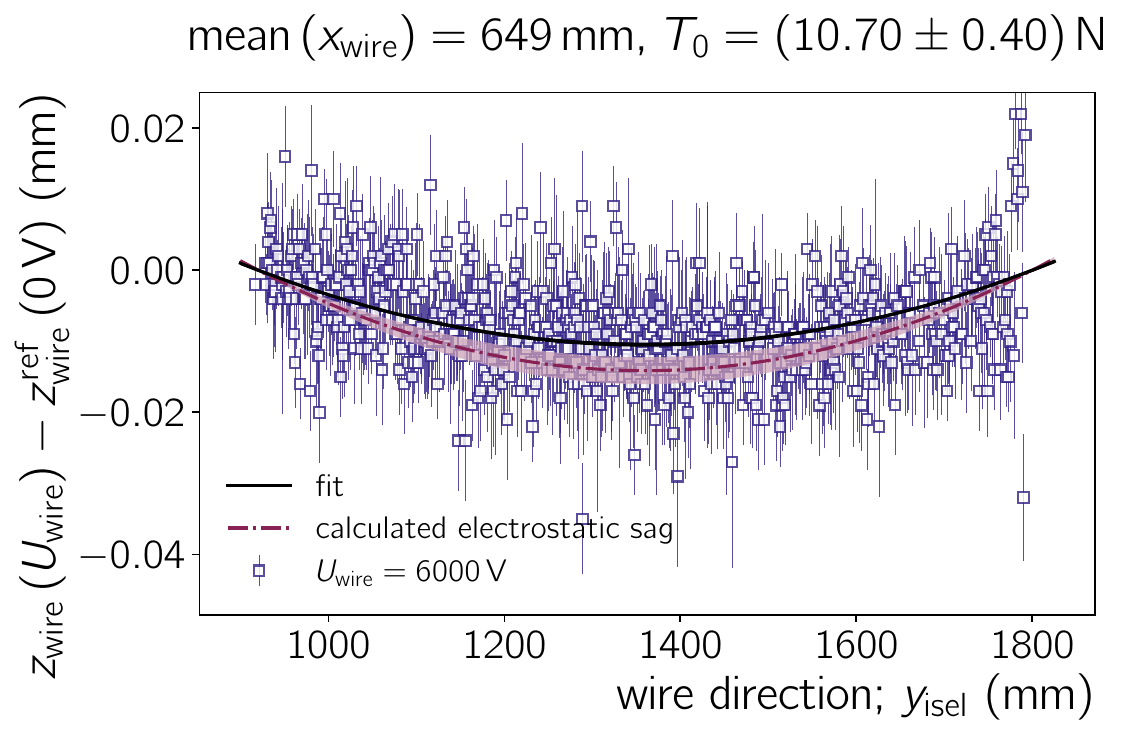}}\\
  \subfloat[]{\label{sec:sagging:test-wires:subsec:sagging-measurement:subsec:electro:fig:meas:5N}\includegraphics[width=0.375\columnwidth]{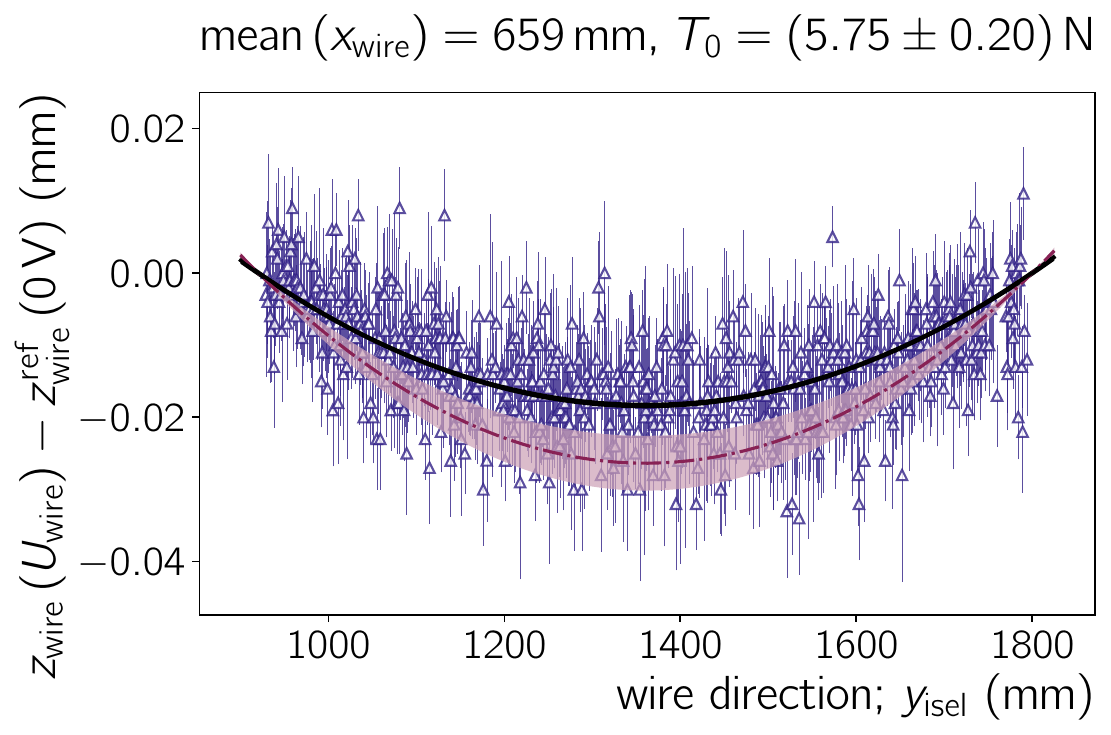}}
  \subfloat[]{\label{sec:sagging:test-wires:subsec:sagging-measurement:subsec:electro:fig:meas:3N}\includegraphics[width=0.375\columnwidth]{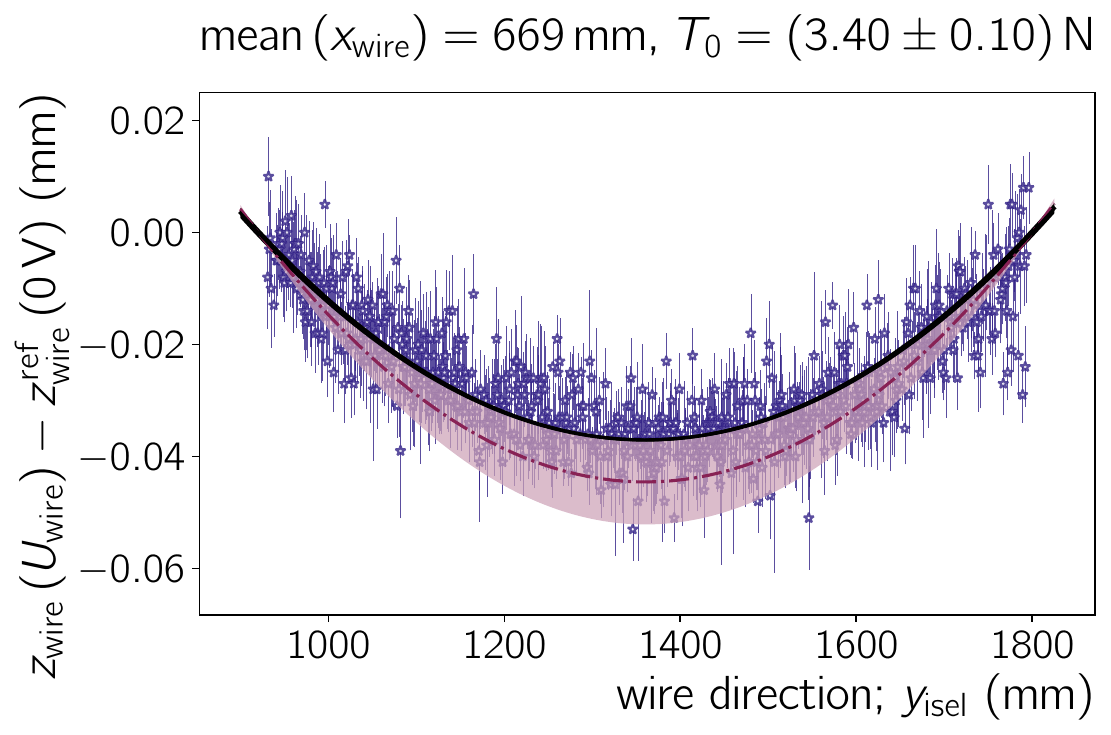}}
  \caption{\label{sec:sagging:test-wires:subsec:sagging-measurement:subsec:electro:fig:explain:meas}Measurement of electrostatic sagging as explained in the text. A fit of \eqnref{sec:sagging:test-wires:subsec:sagging-measurement:eq:sagging:fit} to the data is shown for all wires, together with the calculation of the expected sagging. The legend in \figsubref{sec:sagging:test-wires:subsec:sagging-measurement:subsec:electro:fig:meas:10N:2} applies to all four plots.}
\end{figure}
The sagging of wires can either be described by a catenary or parabolic equation, but the latter is preferred for tensioned wires \cite{Rolandi:2008ujz}. The formula uses the parameters introduced in \eqnref{eq:c:fg:and:fe:all:components:general} and \eqnref{eq:c:fe:all:components}, as well as the parameter $y_\text{isel}^\text{shift}=\SI{1360.1}{\milli\meter}$, which accounts for the shift of the centre of the parabola. 
The general shape difference between the expected and measured curves is due to not yet corrected gantry artifacts. For their removal we can neither use a measurement of the frame rods nor stainless steel ground plane, as their flatness can well be worse than \SI{50}{\micro\meter}. 
One of the four stretched wires could be used as a calibration, if its sagging is taken into account, which makes the measurement model dependent. To do so, a fully corrected elevation profile is fitted with a P10 (black solid line in \figrefbra{sec:sagging:test-wires:subsec:sagging-measurement:subsec:abs:fig:explain:meas:10N:P1corr:sincorr:grav:p10fit}). This curve describes gantry effects as well as the deflection due to gravity, where the latter can be removed by adding the corresponding calculated sagging curve. Subtracting this curve (dashed line in \figrefbra{sec:sagging:test-wires:subsec:sagging-measurement:subsec:abs:fig:explain:meas:10N:P1corr:sincorr:grav:p10fit}) from a wire's elevation profiles should yield the parabola expected for a sagging wire. In \figref{sec:sagging:test-wires:subsec:sagging-measurement:subsec:abs:fig:explain:meas:P1corr:sincorr:p10corr}, the curve is subtracted from all elevation profiles ($\mathcal{C}_\text{P10}$ correction). For reference, the expected sagging for the \SI{0}{\volt} (\textit{calculated gravitational sag} in the figure) and the \SI{6000}{\volt} (\textit{calculated grav. + elec. sag} in the figure) applied HV is plotted as well. Both match in case of the $T_0=\SI{10.4(4)}{\newton}$ wire (\figrefbra{sec:sagging:test-wires:subsec:sagging-measurement:subsec:abs:fig:explain:meas:10N:P1corr:sincorr:p10corr}), which is the basis of the $\mathcal{C}_\text{P10}$ correction.\\
{\indent}The residuals between the measured sagging for $U_\text{wire}=\SI{0}{\volt}$ and the calculated gravitational sagging show a slight decreasing (\figrefbra{sec:sagging:test-wires:subsec:sagging-measurement:subsec:abs:fig:explain:meas:10N:2:P1corr:sincorr:p10corr} and \figrefbra{sec:sagging:test-wires:subsec:sagging-measurement:subsec:abs:fig:explain:meas:5N:P1corr:sincorr:p10corr}) or increasing (\figrefbra{sec:sagging:test-wires:subsec:sagging-measurement:subsec:abs:fig:explain:meas:3N:P1corr:sincorr:p10corr}) slope along the wire direction. The residuals between the calculated sagging due to gravitational and electrostatic forces and the $U_\text{wire}=\SI{6000}{\volt}$ data follow the same trend. 
It is worth to note that after all corrections are applied, there is often an up to $\sim\!\!\SI{25}{\micro\meter}$ height difference at the beginning or end of the wire, with respect to the expected $z_\text{wire}=\SI{0}{\milli\meter}$, \textit{cf.} Figures \ref{sec:sagging:test-wires:subsec:sagging-measurement:subsec:abs:fig:explain:meas:3N:P1corr}, \ref{sec:sagging:test-wires:subsec:sagging-measurement:subsec:abs:fig:explain:meas:3N:P1corr:sincorr:grav}. Based on the residuals between the data after corrections and the predicted sagging by the calculation, we estimate the precision of the corrected and model dependent absolute sagging measurement to be better than \SI{50}{\micro\meter}. Without the use of the model, the precision is better than \SI{200}{\micro\meter}. Both are sufficient for the characterisation of electrodes intended for future TPCs.

\subsubsection*{Sagging Induced by Electrostatic Forces}
\label{sec:sagging:test-wires:subsec:sagging-measurement:subsec:electro}

The combination of all corrections ($\mathcal{C}_\text{P1}$, $\mathcal{C}_\text{sin}$, $\mathcal{C}_\text{P10}$) is based on subtracting a single value at each position along a wire. When subtracting two measurements of a wire in a given $y_\text{isel}$ position -- one reference measurement ($z^{\text{ref}}_{\text{wire}}$) at $U_\text{wire}=\SI{0}{\volt}$ and one at $U_\text{wire}\neq\SI{0}{\volt}$ -- the features induced by the gantry (\textit{cf.} above and in \secrefbra{sec:setup:subsec:gantry:features}) drop out. 
For that reason, a measurement of only the electrostatic sagging is not affected by the \iselSystem{} features. The plots in \figref{sec:sagging:test-wires:subsec:sagging-measurement:subsec:electro:fig:explain:meas} show the difference of the $U_\text{wire}=\SI{6000}{\volt}$ and $U_\text{wire}=\SI{0}{\volt}$ measurements for all four wires. 
The square root of the sum of the squared uncertainties is used as the uncertainty of the subtracted values. 
The parabolic expression introduced in \eqnref{sec:sagging:test-wires:subsec:sagging-measurement:eq:sagging:expectation} allows a factorisation between gravitational and electrostatic sagging, therefore we can fit the electrostatic sagging as:
\begin{align}
  z_\text{wire}\left(y_\text{isel},U_\text{wire}\right) - z^{\text{ref}}_{\text{wire}}\left(y_\text{isel},\SI{0}{\volt}\right) & = p_0 \cdot \left(\left(\frac{l_\text{wire}}{2}\right)^{2}-\left(y_\text{isel}-y_\text{isel}^\text{shift}\right)^2\right)
  \label{sec:sagging:test-wires:subsec:sagging-measurement:eq:sagging:fit}
\end{align}
The only fit parameter is $p_0$ and represents $z_\text{min,E} \cdot \frac{4}{l_\text{wire}^2}$, encoding the maximal sagging for a given wire at the voltage $U_\text{wire}$ as written out in \eqnref{eq:parabola:solution:diff:2dim:maxsagg}. The fit result is shown as well in the various panels of \figref{sec:sagging:test-wires:subsec:sagging-measurement:subsec:electro:fig:explain:meas} together with the calculated electrostatic sagging. The calculation and its uncertainty overestimate the sagging by \SI{5}{\micro\meter} to \SI{10}{\micro\meter}, but the calculation assumes an ideal electric field of a wire above a grounded plane. However, in the discussed measurements the electric fields are reduced by edge effects from the frame. The first row of \figref{sec:sagging:test-wires:subsec:sagging-measurement:subsec:electro:fig:explain:meas} also illustrates that  deflections due to the electric field smaller than $\SI{20}{\micro\meter}$ can be measured. This is well below the  requirement for a sagging measurement defined at the beginning of this section.

\section{Optical Analysis of the XENON1T Cathode}
\label{sec:optical:analysis:1T}

Having established the performance of GRANITE when measuring wire tension and sagging for well known test wires, we move now to the assay of a full electrode, the cathode of the XENON1T dual-phase TPC \cite{XENON:2017lvq}. Optical inspection complements the mechanical characterisation, as it allows to identify local defects, which may impact the performance of the electrode inside the detector. When the experiment was upgraded to become XENONnT, the XENON1T cathode became available and it served as first major use-case for GRANITE's assay-by-imaging capabilities.\\
{\indent}This electrode is a grid with 124 parallel wires (\SI{7.5}{\milli\meter} pitch) and a total diameter of \SI{95}{\centi\meter}. The wires are gold plated stainless-steel with a diameter of \SI{216}{\micro\meter}. From the operation of XENON1T, regions with an elevated rate of single electron (SE) emission were already known before the optical inspection. Scanning the cathode is thus motivated by the possibility to correlate damages observed during the inspection with regions of enhanced SE emission inside the detector.

\subsection{Measurement Procedure}
\label{sec:optical:analysis:1T:subsec:meas}

\begin{figure}[t]
  \centering
  \subfloat[]{\label{sec:optical:analysis:1T:subsec:meas:fig:1t_cathode_inspection:map}
    \includegraphics[width=0.56\columnwidth]{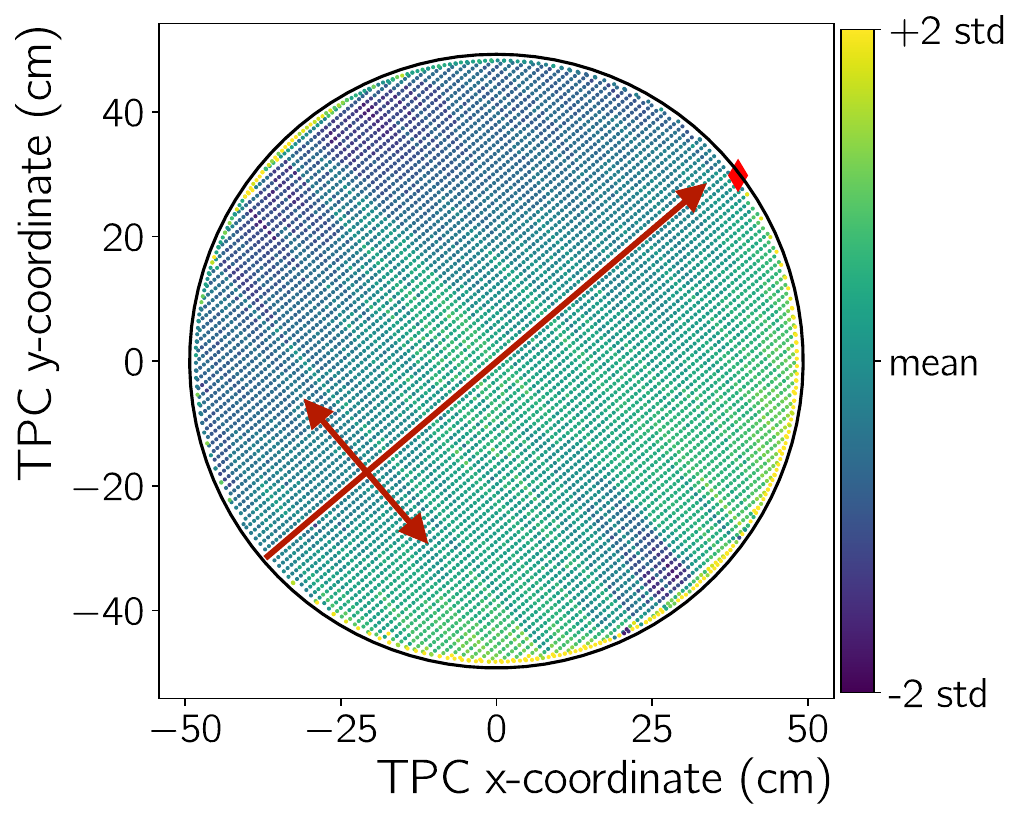}}
  \subfloat[]{\label{sec:optical:analysis:1T:subsec:meas:fig:1t_cathode_inspection:wires}
    \includegraphics[width=0.48\columnwidth,angle=90]{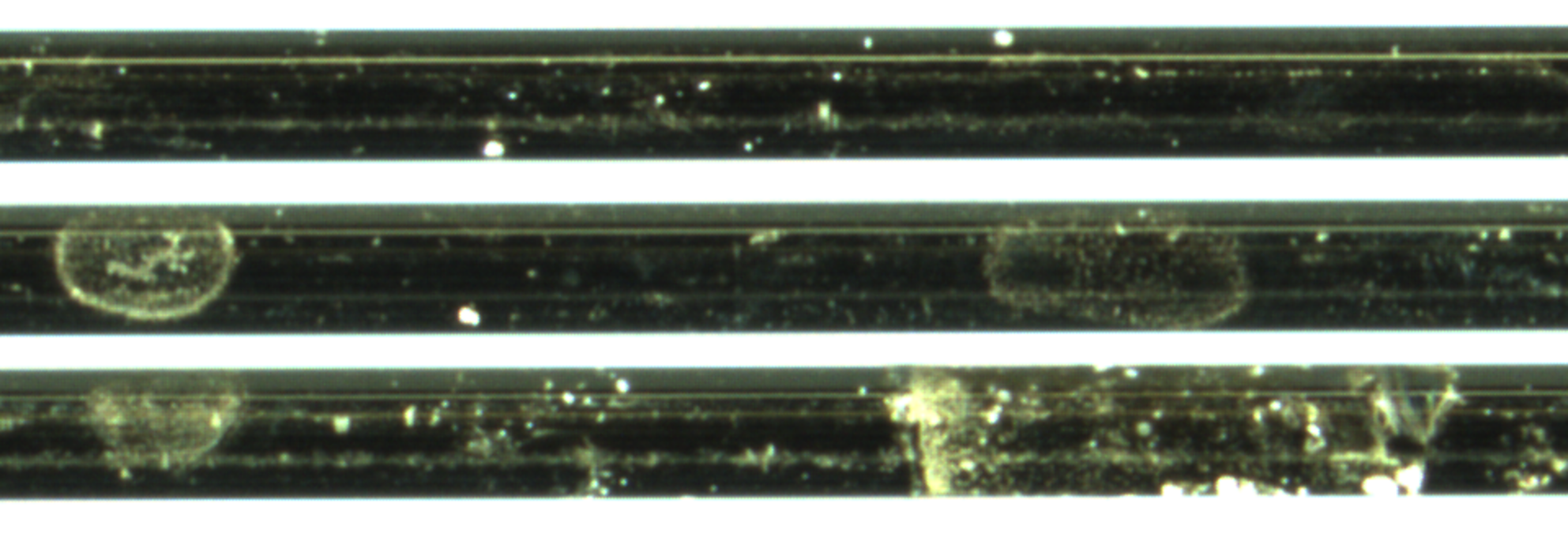}}
  \caption{\label{sec:optical:analysis:1T:subsec:meas:fig:fig::1t_cathode_inspection}\figsubref{sec:optical:analysis:1T:subsec:meas:fig:1t_cathode_inspection:map} Illustration of the automated scanning procedure of the XENON1T cathode: The scan starts at the outer most wire in the bottom left quadrant. Every wire is scanned beginning at its centre, moving in both directions as indicated with the red arrow. Then, the \iselSystem{} moves the camera along the longer red arrow by one wire pitch to scan the next wire. The colour-bar encodes the $z_\text{isel}$ position for the respective wire, normalized to its average $z_\text{isel}$ over the full wire. This height for recording each image is determined by the autofocus algorithm explained in \cite{Wenz:2023qzq}, and shows sagging together with the \iselSystem{} induced height offset. The red diamond marks the location of the HV connection within the experiment. \figsubref{sec:optical:analysis:1T:subsec:meas:fig:1t_cathode_inspection:wires} Different wire defects found on the grid.}
\end{figure}
For the assay, the cathode is placed on PTFE spacers on top of the granite table. Then, the \textit{Basler acA4600-7gc} camera equipped with the \textit{TC23016} telecentric lens is moved across the wire grid in the manner illustrated in \figref{sec:optical:analysis:1T:subsec:meas:fig:1t_cathode_inspection:map}. Images are recorded along every wire. The lens projects non-equidistant objects as same-sized objects onto the image plane. Wires appear in images thus as flat objects, aiding defect recognition algorithms latter on. Multiple lighting configurations were tested and dark field illumination was found optimal for defect detection. 
At least three images are taken at each position along each wire to establish the best focus. To do so, a passive autofocus algorithm has been developed for GRANITE, moving the gantry in $z_\text{isel}$ to change the camera height when a wire is out of focus. \Figref{sec:optical:analysis:1T:subsec:meas:fig:1t_cathode_inspection:map} illustrates the height differences throughout the scan, which are caused by the wire sagging and the gantry features discussed in \secref{sec:setup:subsec:gantry:features} and \secref{sec:sagging:test-wires}. Further details on the selection of the telecentric lens as well as on the workings of \textit{e.g.} the autofocus can be found in \cite{Wenz:2023qzq}.\\
{\indent}Throughout the fully automatised XENON1T cathode scan, $8\,400$ images (\SI{75}{GB} in total) were written to disk over a scan duration of 60 hours. 
Given the dark field illumination a perfectly smooth surface would appear as dark in a camera image. Deformations or contaminations become visible as they reflect light onto the camera sensor. Therefore, the greater part of the wires in \figref{sec:optical:analysis:1T:subsec:meas:fig:1t_cathode_inspection:wires} appear as black. Several features are revealed as bright spots or larger areas of discolourations. 
Other features visible are salt remnants of the cleaning procedure (middle wire image), as well as large-scale defects (right wire image).

\subsection{Image Classification with an Autodencoder}
\label{sec:optical:analysis:1T:subsec:imageana}

In order for a quantitative analysis of defects or other ``atypical'' regions on the wires of the XENON1T cathode grid 
these regions need to be identified. Searching all images by eye is not feasible, therefore an autoencoder was used. An autoencoder is an unsupervised machine learning algorithm, which has been historically employed for data compression, but is now widely relied on to detect anomalies in input data \cite{BELIS2024100091}. The autoencoder classification distinguishes ``least-concern'' (\textit{i.e.} predominately dark) from ``greatest-concern'' (\textit{i.e.} many bright features visible) images of wire segments. This is done by teaching a convolutional neural network (CNN) with \textit{least-concern} image data to encode and reconstruct the given sample image. The difference between the input image and the encoded-and-decoded image is measured by the mean square error (MSE), which 
is a common loss function for this application. It measures the average squared difference between each pixel value in the original input image and the decoded output image.\\
\begin{figure}
  \subfloat[]{\label{sec:optical:analysis:1T:subsec:imageana:fig:trainingandcalibration:loss:v:epoch}
    \includegraphics[width=0.49\columnwidth,trim=0 5 0 0,clip=true]{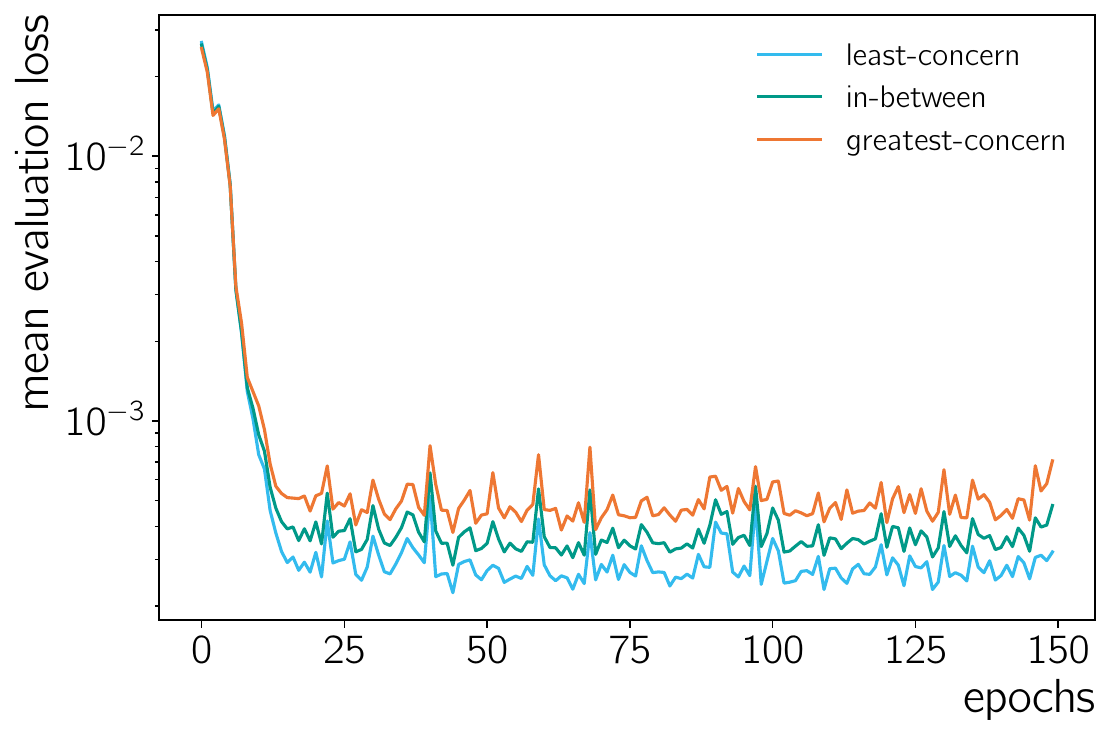}}
  \subfloat[]{\label{sec:optical:analysis:1T:subsec:imageana:fig:trainingandcalibration:hist:good:v:bad}
    \includegraphics[width=0.49\columnwidth,trim=0 5 0 0,clip=true]{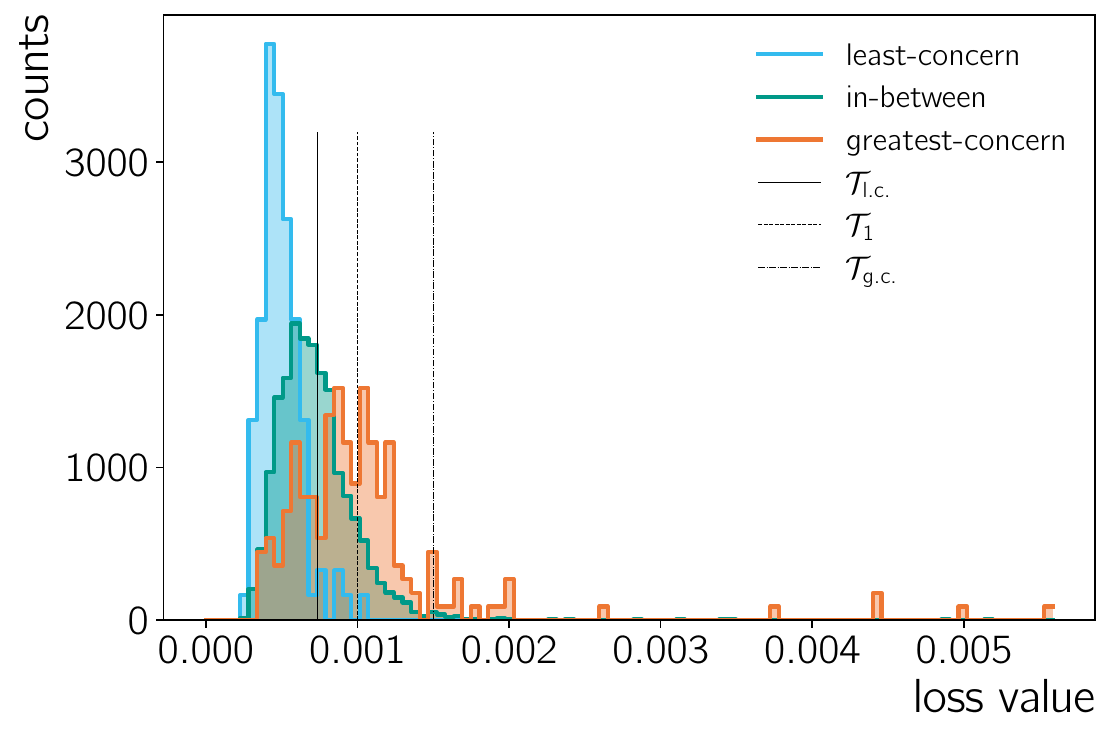}}\\[0.1cm]
  \subfloat[]{\label{sec:optical:analysis:1T:subsec:imageana:fig:trainingandcalibration:good:vs:bad:vs:loss}
    \includegraphics[width=0.49\columnwidth,trim=0 5 0 0,clip=true]{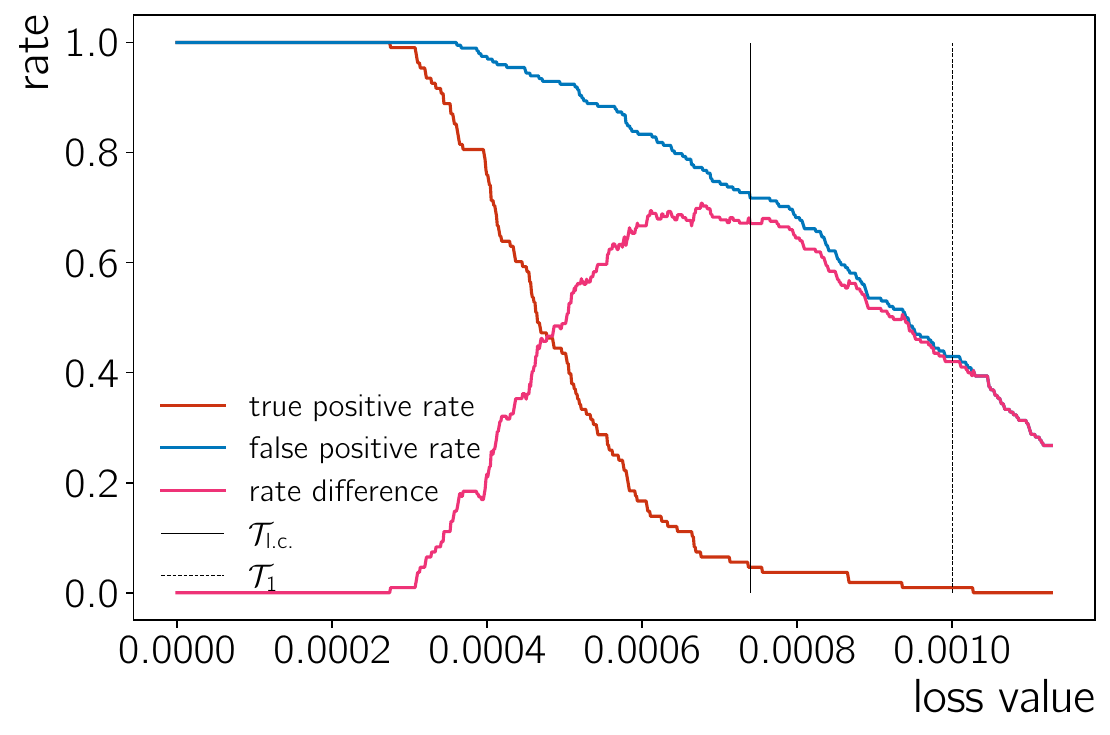}}
  \subfloat[]{\label{sec:optical:analysis:1T:subsec:imageana:fig:trainingandcalibration:other:vs:bad:vs:loss}
    \includegraphics[width=0.49\columnwidth,trim=0 5 0 0,clip=true]{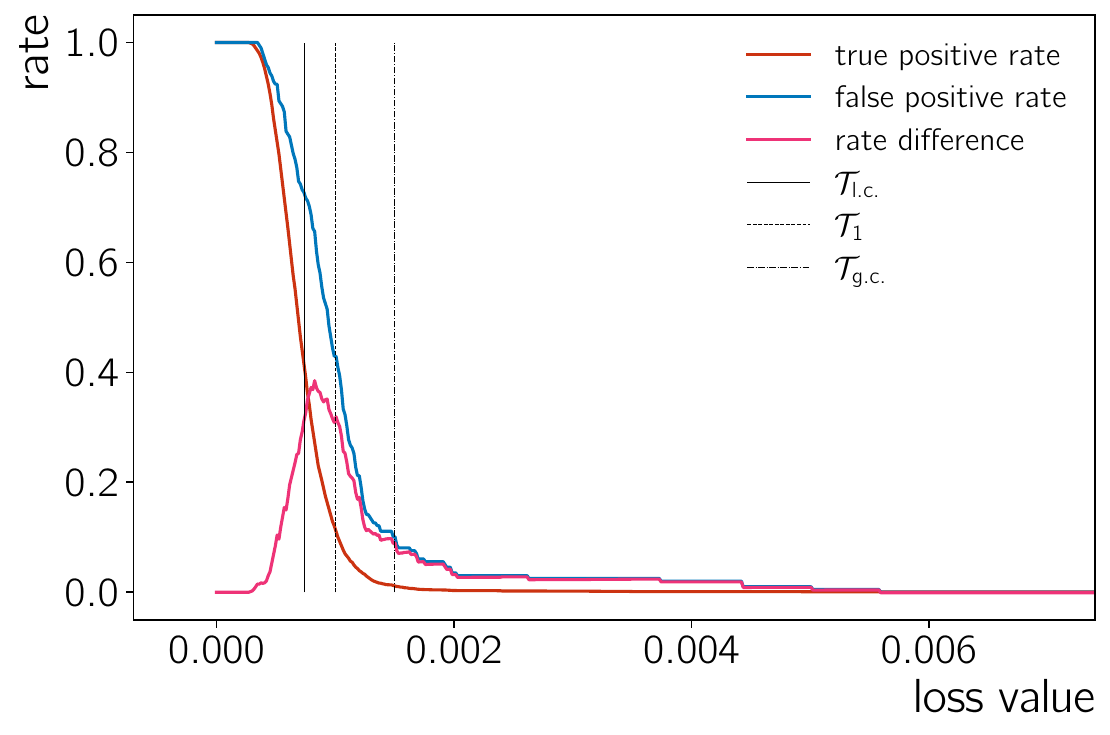}}
  \caption{\label{sec:optical:analysis:1T:subsec:imageana:fig:trainingandcalibration}\figsubref{sec:optical:analysis:1T:subsec:imageana:fig:trainingandcalibration:loss:v:epoch} Average loss value for \textit{least-concern} and \textit{greatest-concern} images of wire segments, and such images, which fit neither classification well (\textit{in-between}), after each training epoch using unseen samples. The clear separation after a couple of epochs already indicates the validity of the approach -- \textit{least-concern} samples should in general have a lower loss value than images showing some anomaly. \figsubref{sec:optical:analysis:1T:subsec:imageana:fig:trainingandcalibration:hist:good:v:bad} Loss value distribution for passing previously hand selected calibration images through the trained autoencoder. 
  False-positive-rate and true-positive-rate as well as their difference as function of the evaluation loss for running the classification on \figsubref{sec:optical:analysis:1T:subsec:imageana:fig:trainingandcalibration:good:vs:bad:vs:loss} \textit{least-concern} and \textit{greatest-concern} images as well as \figsubref{sec:optical:analysis:1T:subsec:imageana:fig:trainingandcalibration:other:vs:bad:vs:loss} \textit{in-between} and \textit{greatest-concern} images. The different black lines in \figsubref{sec:optical:analysis:1T:subsec:imageana:fig:trainingandcalibration:hist:good:v:bad}, \figsubref{sec:optical:analysis:1T:subsec:imageana:fig:trainingandcalibration:good:vs:bad:vs:loss}, \figsubref{sec:optical:analysis:1T:subsec:imageana:fig:trainingandcalibration:other:vs:bad:vs:loss} are thresholds, to select between the different classes: losses smaller than $\mathcal{T}_{l.c.}$ (greater than $\mathcal{T}_{g.c.}$) select predominately \textit{least-concern} (\textit{greatest-concern}) wire segments, with only little contamination by segments in the \textit{greatest-concern} (\textit{least-concern} and \textit{in-between}) category. Loss values larger than the $\mathcal{T}_{1}$ threshold select $\sim\!\!\SI{50}{\%}$ of \textit{greatest-concern} images, with $\sim\!\!\SI{10}{\%}$ contamination from the \textit{in-between} category.}
\end{figure}
{\indent}The autoencoder has an under-complete architecture using 2D convolutional layers with (8,4,4,8,3) filter employing 7$\times$7 kernels and max-pooling layers (encoding side) or upsampling layers (decoding side) in between. 
Hand scanned image data deemed \textit{least-concern} from wires 70 to 125 -- about 2000 images -- have been selected as training data and are fed to the CNN. Doing so, \SI{10}{\%} of this data are held back for validation after each training epoch. 
Before feeding images into the CNN, they are resized to $536\times400$ pixel, the colour channels are rescaled to values between 0 and 1, the image is randomly rotated by an angle between $0\,^{\circ}$ and $45\,^{\circ}$, and Gaussian noise is applied with a standard deviation of 0.3. These steps are necessary to avoid over-fitting due to the small training set. 
The output is evaluated against the input using the previously mentioned MSE. 
The total number of trainable parameters is 6299, which is still in the same order of magnitude as the available training set.\\[0.2cm]
\begin{figure}
  \subfloat[]{\label{sec:optical:analysis:1T:subsec:fullgrid:fig:voxels:min=0.00000-max=0.00074}
    \includegraphics[width=0.49\columnwidth]{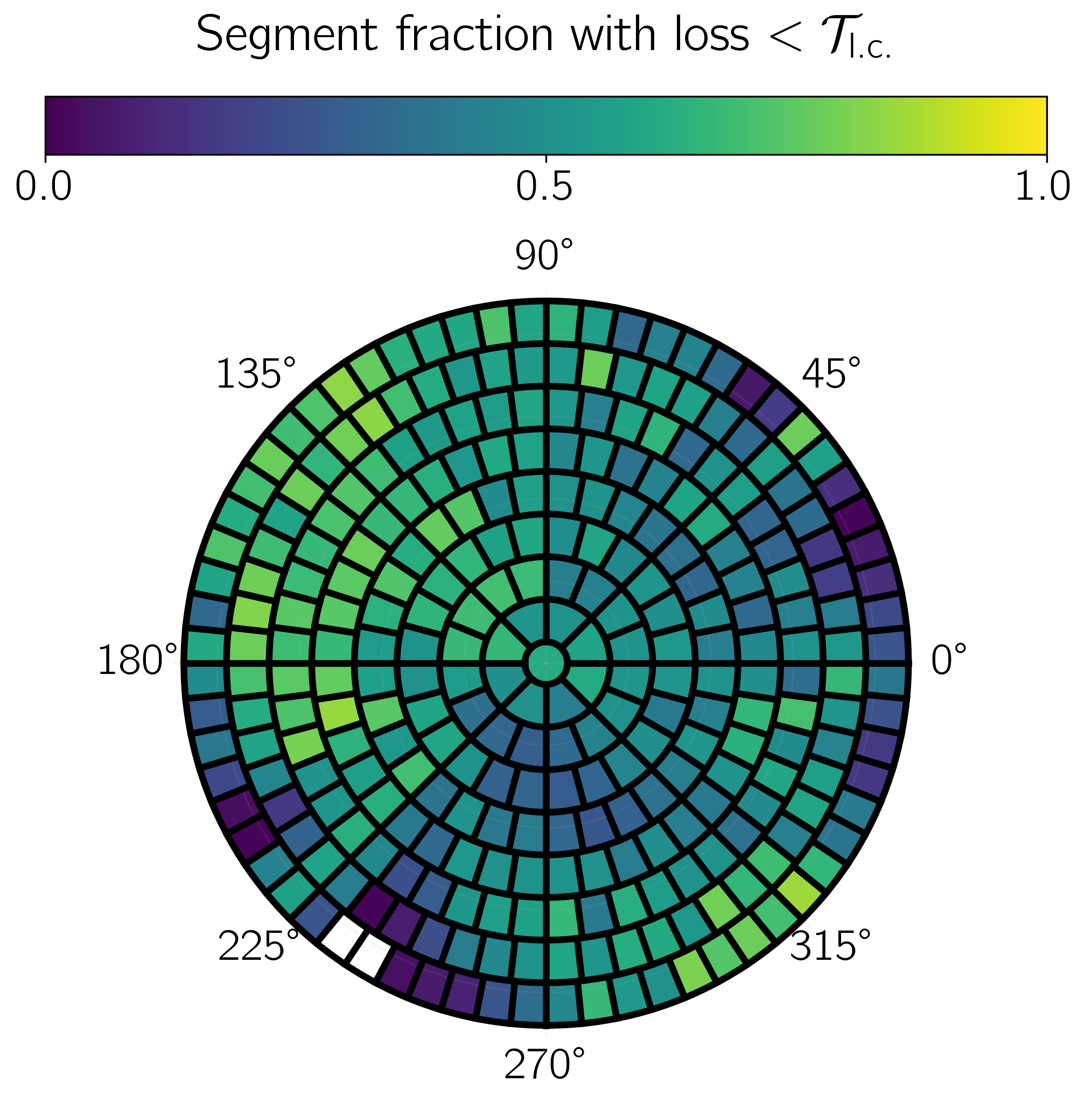}}
  \subfloat[]{\label{sec:optical:analysis:1T:subsec:fullgrid:fig:voxels:min=0.00074-max=0.00100}
    \includegraphics[width=0.49\columnwidth]{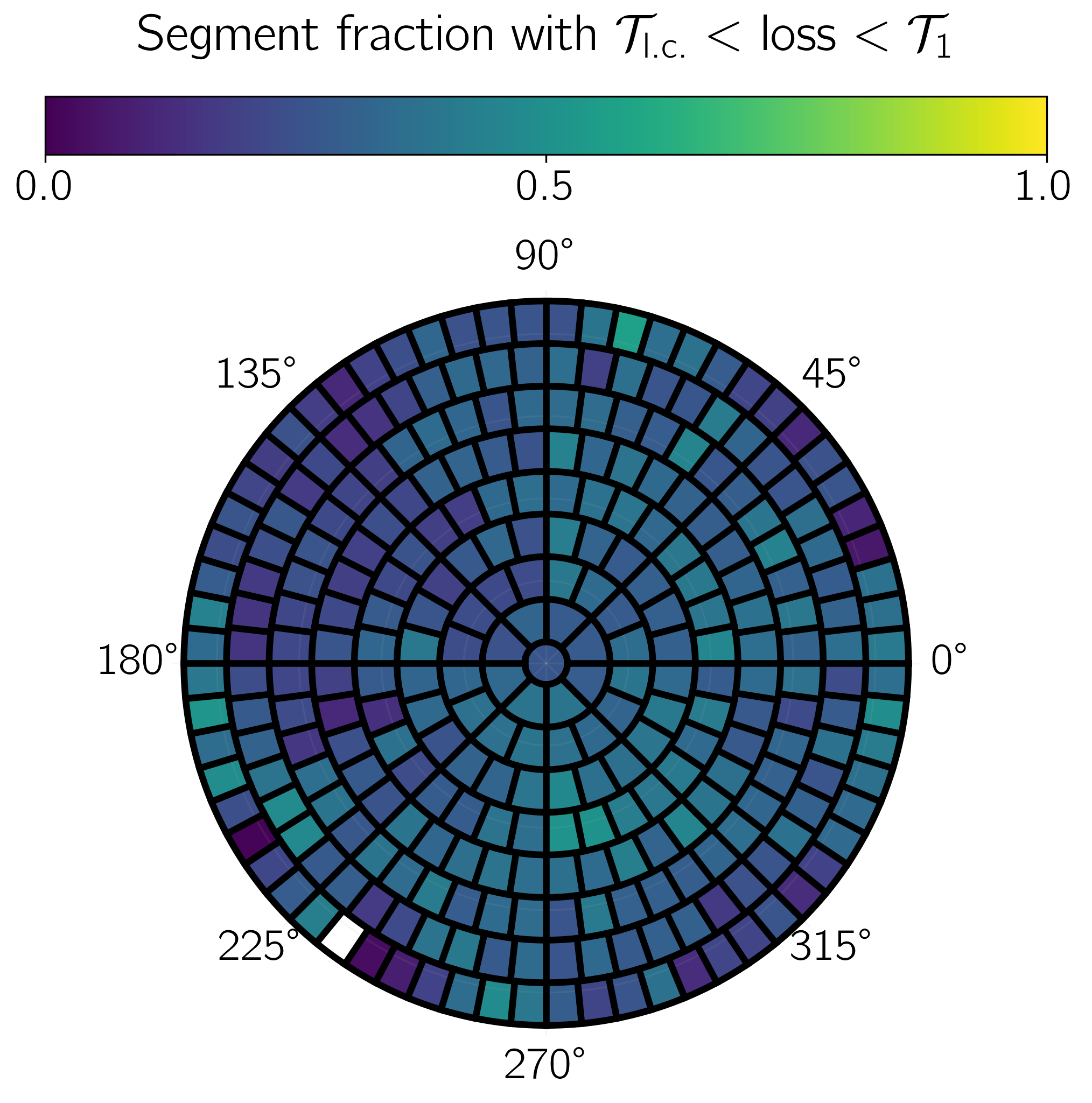}}\\[0.1cm]
  \subfloat[]{\label{sec:optical:analysis:1T:subsec:fullgrid:fig:voxels:min=0.00100-max=0.00150}
    \includegraphics[width=0.49\columnwidth]{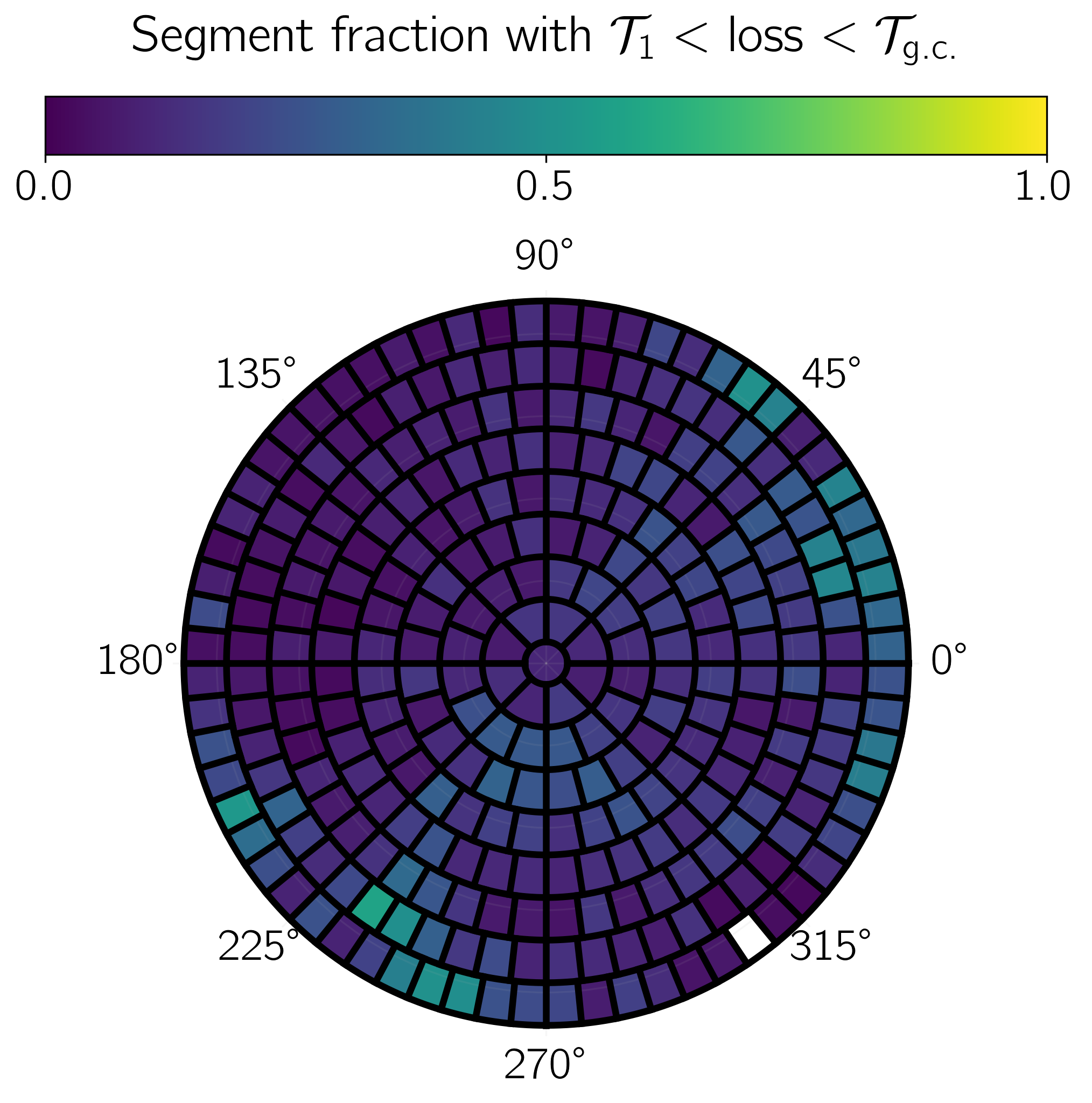}}
  \subfloat[]{\label{sec:optical:analysis:1T:subsec:fullgrid:fig:voxels:min=0.00150-max=1.00000}
    \includegraphics[width=0.49\columnwidth]{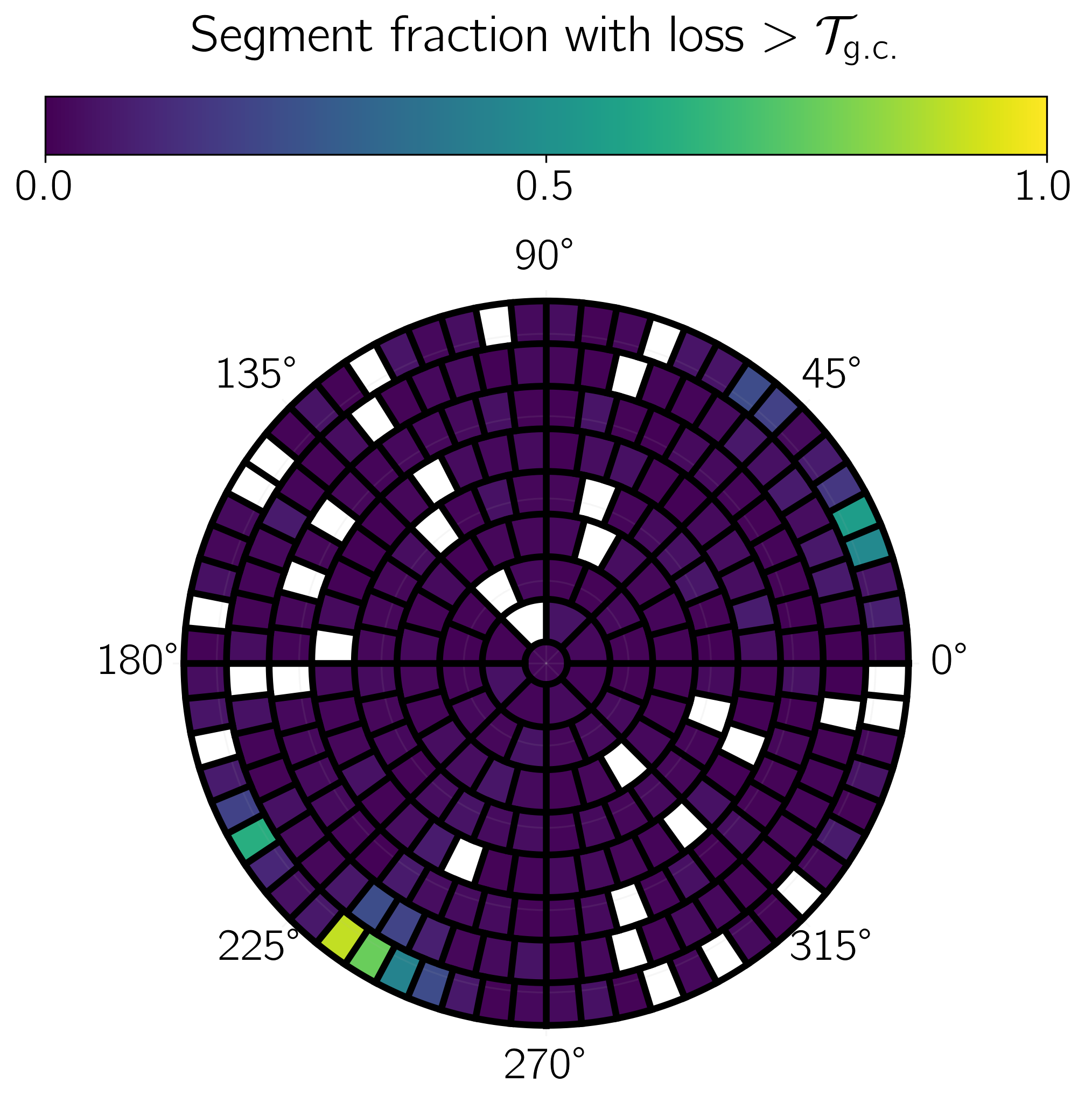}}
  \caption{\label{sec:optical:analysis:1T:subsec:fullgrid:fig:voxels:2}Binned loss maps based on feeding images within the inner \SI{44}{\centi\meter} radius of the XENON1T cathode grid through the autoencoder and comparing the result to the input data. Each bin contains information of approximately 200 wire segments. The colour scale encode what fraction of wire segments within the bin fall into the loss-threshold range indicated above the colour bar. The following classes occur in the corresponding ranges:  
  \figsubref{sec:optical:analysis:1T:subsec:fullgrid:fig:voxels:min=0.00000-max=0.00074} 
  \SI{90}{\%} \textit{least-concern} segments, \SI{50}{\%} \textit{in-between} segments, \SI{30}{\%} \textit{greatest-concern} segments; \figsubref{sec:optical:analysis:1T:subsec:fullgrid:fig:voxels:min=0.00074-max=0.00100} the same fraction of \textit{in-between} and \textit{greatest-concern} segments, with little contamination from \textit{least-concern} segments; \figsubref{sec:optical:analysis:1T:subsec:fullgrid:fig:voxels:min=0.00100-max=0.00150} no \textit{least-concern} segments, more \textit{greatest-concern} than \textit{in-between} segments;
  \figsubref{sec:optical:analysis:1T:subsec:fullgrid:fig:voxels:min=0.00150-max=1.00000} exclusively \textit{greatest-concern} segments. Percent values given are only approximate and are based on the calibration data set. Bins with no wire segment images matching the conditions on their loss values are not filled.}
\end{figure}
{\indent}During training, all images from wires 63 to 69 are used to calibrate the loss value for the final classification of wire segment images. Again, the images are hand scanned and group into \textit{least-concern}, \textit{greatest-concern}, and everything \textit{in-between} -- depending on the abundance of possible anomalies in these images. When feeding this data through the autoencoder, it is apparent that the average loss value between the three classes separates after few epochs, and that \textit{least-concern} and \textit{greatest-concern} exhibit the smallest and largest loss values (\figrefbra{sec:optical:analysis:1T:subsec:imageana:fig:trainingandcalibration:loss:v:epoch}). The latter fact, together with the clear separation of the two ``extreme cases'' in \figref{sec:optical:analysis:1T:subsec:imageana:fig:trainingandcalibration:hist:good:v:bad}, confirms that our classification approach is sensible. \Figref{sec:optical:analysis:1T:subsec:imageana:fig:trainingandcalibration:good:vs:bad:vs:loss} shows furthermore, that a robust anomaly classification between clean (\textit{least-concern}) and problematic wire segments (\textit{greatest-concern}) can be achieved with a loss threshold of $\mathcal{T}_{l.c.}=7.4\times10^{-4}$. However, there is a large fraction of images fitting in neither category clearly. This \textit{in-between} population has a large tail towards higher loss values (\figrefbra{sec:optical:analysis:1T:subsec:imageana:fig:trainingandcalibration:other:vs:bad:vs:loss} and \figrefbra{sec:optical:analysis:1T:subsec:imageana:fig:trainingandcalibration:good:vs:bad:vs:loss}). To select a clean sample of the \textit{greatest-concern} class, a loss threshold of $\mathcal{T}_{g.c.}=1.5\times10^{-3}$ is deemed necessary. For loss values $\geq\mathcal{T}_{g.c.}$, almost exclusively \textit{greatest-concern} segments remain. A third threshold of $\mathcal{T}_{1}=1.0\times10^{-3}$ is defined. Loss values above that, select almost no wire segments form the \textit{least-concern} class and about $\SI{50}{\%}$ of \textit{greatest-concern} images, with approximate $\SI{10}{\%}$ contamination from the \textit{in-between} category.

\subsection{XENON1T Cathode Grid Anomaly Discussion}
\label{sec:optical:analysis:1T:subsec:fullgrid}

Finally, the trained autoencoder is fed with the full set of images of the XENON1T cathode grid. The obtained loss values (\figrefbra{sec:optical:analysis:1T:subsec:fullgrid:fig:voxels:2}) are shown as a two-dimensional histogram, using a polar binning based on the XENON1T voxelization during the experiment's $^{37}\!\text{Ar}$ analysis \cite{XENON:2022ivg}. 
The figure with all its subplots demonstrates that segments with high and low loss values are evenly distributed over the whole cathode. The bins on edges of the gird appear as exceptions to that statement, despite including only wire images within a inner, \SI{44}{\centi\meter} radius, disk. However, at these edges the light conditions can be different, due to shade from the cathode's frame. Therefore it can not be concluded that the wire quality is significantly worse in these regions.\\
{\indent}The main result is thus, that the wires of this cathode grid are full of anomalous spots. This fact makes a direct correlation with the observed SE emission in XENON1T not feasible. Furthermore, the gate electrode is at least as likely to have contributed to the SE emission inside the experiment as the cathode. By optical inspection alone, and given the anomaly rate, there is no possibility of telling which of these anomalies may lead to SE emission in the detector or may affect detector performance in a different way. A direct, table top, test of field emission in an electric field is needed, if the nature of potential defects visible in images is to be better understood. A first proposal of such studies can be found in \cite{Wenz:2023qzq} and first work on the topic in \cite{Mitra:2023}.

\section{Conclusions and Outlook}
\label{sec:conclusions:outlook}

To analyse whether large-area electrodes meet the stringent requirements for employment in future dual-phase TPCs, we have developed the GRANITE (Granular Robotic Assay for Novel Integrated TPC Electrodes) setup, incorporating laser-distance sensors, a high resolution camera and a confocal microscope into a moveable gantry robot mounted on top of a granite table. All measurements can be fully automated using custom software.\\
{\indent}First, we demonstrate the setup's ability to perform high-precision mechanical characterization of electrode wires. With the laser distance sensors the setup can measure sagging directly, and independently measure the wire tension of parallel-wire electrodes. For the relative electrostatic sagging measurements a precision of \SI{20}{\micro\meter} is achieved. While absolute wire deflection measurements are limited by gantry-induced systematics, gravitational sagging can still be measured to values as low as \SI{200}{\micro\meter}. With the aid of reference measurements and models \SI{50}{\micro\meter} can be reached. Overall, the measurement precision is sufficient to characterize electrodes for future detectors.\\
{\indent}Second, GRANITE is used to perform an optical scan of all wires of the XENON1T cathode grid (\SI{95}{\centi\meter} diameter, 125 parallel wires). The images resulting from a \SI{60}{h} scan with a high resolution camera are then fed into an autoencoder, which is trained to classify between images with spotless wire segments and such exhibiting anomalies. After training, the autoencoder can reliably sort wire images into these classes. A global analysis of the full grid reveals that most images show some anomalies. Therefore, it is not possible to correlate features on the grid with known XENON1T detector performance parameters as \textit{e.g.} known locations of SE emission. However, the test demonstrates the fully automated optical scan of a real electrode.\\
{\indent}Work is ongoing to integrate laser-distance sensors as well as high resolution cameras into a new set-up, based on a collaborative robot (``cobot''), which will have the reach to scan up to \SI{3}{\meter} diameter electrodes. This will allow for a mechanical characterisation as well as high-resolution imaging of XLZD-sized meshes or grids.\\
{\indent}Furthermore, the observed prevalence of anomalies stresses the need for a wire-by-wire study of local field emission in combination with high resolution imaging. Such a study may shed light on what kind of anomalies lead to problems in a future detector, for example by emitting SEs and contributing to the background in a corresponding experiment. We report on this work, using GRANITE once more, in \cite{Smitra2025}.

\section*{Acknowledgements}

This work has been supported by the Cluster of Excellence “Precision Physics, Fundamental Interactions, and Structure of Matter” (PRISMA$^{+}$ EXC 2118/1) funded by the German Research Foundation (DFG) within the German Excellence Strategy (Project ID 390831469). The authors thank L. F. Deibert for her help during the data taking. Further support was received by BMBF under grant number 05A20UM1. D. Wenz was supported by the German Academic Scholarship Foundation.


\bibliographystyle{JHEP}
\bibliography{notes.bib}

\end{document}